\RequirePackage[2024-06-01]{latexrelease}

\documentclass[aps,prx,twocolumn,superscriptaddress,floatfix,longbibliography]{revtex4-2}

\usepackage{blindtext}
\usepackage{amsmath}

\usepackage{bm}
\usepackage{gensymb}
\usepackage[normalem]{ulem}
\usepackage{euscript}
\usepackage{graphicx}
\usepackage{xspace}
\usepackage{xfrac}
\usepackage{enumerate}
\usepackage{braket}
\usepackage{float,fancyvrb}
\usepackage[T1]{fontenc}
\usepackage{lmodern}
\usepackage{tabularx,array}

\graphicspath{{./Figures/}}
\usepackage{wrapfig}
\usepackage{caption}
\usepackage{scrextend}
\usepackage{url}
\usepackage{amssymb}
\usepackage{xcolor}
\usepackage{comment}
\usepackage{ragged2e}

\usepackage[utf8]{inputenc}
\usepackage[subrefformat=parens,labelformat=parens,justification=raggedright,singlelinecheck=false]{subcaption}

\usepackage[pdftex,hyperfigures,breaklinks,colorlinks,citecolor=black,linkcolor=black]{hyperref}

\def\be{\begin{equation}}
\def\ee{\end{equation}}
\def\beq{\begin{eqnarray}}
\def\eeq{\end{eqnarray}}
\def\ba#1{\begin{array}{#1}}
\def\ea{\end{array}}
\def\bn{\begin{enumerate}}
\def\en{\end{enumerate}}

\renewcommand{\thesection}{\arabic{section}}

\begin{document}
\linespread{1.1}\selectfont

\title{Variational Iterative Rotation Algorithm: Combinatorial Optimization with Classical Kicked Tops}

\author{Flaviano Morone}
\affiliation{Center for Quantum Phenomena, Department of Physics, New York University, New York, NY 10003 USA}

\author{Andrew D. Kent}
\affiliation{Center for Quantum Phenomena, Department of Physics, New York University, New York, NY 10003 USA}

\author{Dries Sels}
\affiliation{Department of Physics, Boston University, 590 Commonwealth Ave., Boston, Massachusetts 02215, USA} 
\affiliation{Center for Computational Quantum Physics, Flatiron Institute, New York, NY, USA}

\begin{abstract}
We investigate a classical formulation of the 
Quantum Approximate Optimization Algorithm (QAOA), 
realized as a Hamiltonian dynamical system of 
classical kicked tops, which we call the Variational 
Iterative Rotation Algorithm (VIRAL). 
The protocol is fully variational, in that the 
initial tilt angle of the spins and the longitudinal 
and transverse rotation angles of each of the 
$p$ layers are optimized by backpropagation 
through the deterministic dynamics. On the 
canonical Sherrington-Kirkpatrick (SK) spin glass 
and on Max-Cut for random 3-regular graphs, 
VIRAL systematically outperforms QAOA, 
with the SK energy density converging to the 
Parisi ground state value and the cut fraction 
to the 2RSB upper bound, both  linearly in $1/p$.
The optimized protocol admits a clear dynamical 
interpretation, whereby a pitchfork bifurcation 
of the Floquet map destabilizes the equatorial 
fixed point and adiabatically transports 
the spins toward the polar Ising configurations 
that encode the solution. 
These results indicate that the effectiveness 
of QAOA-like protocols is rooted in their 
iterative rotation structure, and that a 
classical implementation of this structure 
outperforms its quantum counterpart. 
To explain this advantage, we exactly reduce 
the many-body classical evolution to an effective 
Landau-Lifshitz dynamics of a single spin in a 
self-consistent stochastic magnetic field. In 
this picture, the temporal covariance matrix of 
the effective field is nearly rank-one, so that 
a single noise mode governs the stochastic dynamics 
and the system remains highly controllable. 
Quantum fluctuations, by contrast, increase the 
rank of the noise covariance of the effective 
quantum model, hampering control and slowing 
convergence to the solution. 
Finally, we propose nanometer-scale magnetic 
tunnel junctions, in which spin rotations are 
naturally realized by magnetic fields and spin 
torques, as a physical platform for implementing 
VIRAL.
\end{abstract}

\maketitle

\onecolumngrid

\section{Introduction}
\label{sec:introduction}
Hard combinatorial optimization problems can often 
be reformulated as finding the ground state of an 
Ising spin-glass system~\cite{lucas2014ising}, a 
perspective that has brought together statistical 
physics, computer science, and electrical engineering~\cite{mezard2009information}, and 
motivated both classical approaches, such as Ising machines~\cite{mohseni2022ising, grimaldi2023evaluating}, 
and quantum algorithms~\cite{carleo2017solving, cerezo2021variational, carleo2019machine}. 
Among the latter, variational quantum algorithms 
have attracted particular attention, as they can 
be realized on noisy intermediate-scale quantum 
devices while having the potential for quantum 
advantage. A prominent example is the Quantum 
Approximate Optimization Algorithm (QAOA)~\cite{farhi2014quantum, farhi2020quantumapproximateoptimizationalgorithm, farhi2022, basso22, cerezo2021variational}, which seeks 
low-energy states of interacting quantum spins 
through a sequence of alternating unitary rotations. 
The performance of QAOA is attributed to quantum 
mechanical effects, yet this attribution remains 
unproven, sparking a growing debate on whether 
quantum resources are truly essential for 
optimization, or whether comparable performance 
can be achieved with classical dynamical systems~\cite{aadit2022massively,medvidovic2021classical}. 
Recent works have begun to address this question 
by exploring classical variational architectures 
that rival or outperform QAOA on benchmark problems~\cite{weitz2025, camsari2025, misra2023mean}. 
In particular, the mean-field Approximate 
Optimization Algorithm~\cite{misra2023mean} 
replaces the unitary evolution of the QAOA 
with the corresponding mean-field spin dynamics 
and already surpasses the QAOA on most instances 
of the Sherrington-Kirkpatrick (SK) and 
number-partitioning problems, even though its 
control schedule is fixed a priori rather than 
variationally optimized. 
A related open question concerns the role of 
stochasticity in the optimization. Noise is 
known to be beneficial in classical settings, 
as exemplified by simulated annealing~\cite{kirkpatrick1983optimization} 
and stochastic resonance~\cite{benzi1981}, but 
whether the quantum fluctuations at the core 
of QAOA constitute a computational resource or 
a detrimental source of noise remains unresolved. 

In this work, we address both questions by 
formulating and training the Variational Iterative 
Rotation Algorithm (VIRAL), a purely classical 
algorithm built on a dynamical system of three-component 
unit vectors (kicked tops)~\cite{haake1987classical}. 
While the underlying alternating dynamics coincides 
with the mean-field limit of the QAOA 
circuit~\cite{misra2023mean}, our protocol is fully 
variational. The deterministic evolution of the 
spins is differentiable with respect to all control 
parameters, so the initial tilt angle and the $2p$ 
rotation angles of a depth-$p$ circuit can be trained 
by backpropagating the gradients of the final energy 
through the layered dynamics, in direct analogy with 
the training of deep neural networks. The resulting 
protocol systematically outperforms QAOA on the SK 
spin-glass problem~\cite{sherrington1975solvable}, 
where the energy density converges linearly in $1/p$ 
to the Parisi ground state value, as well as on 
Max-Cut for random 3-regular graphs. 
Moreover, since the entire algorithm is classical, 
the optimization can be carried out on large 
instances and at circuit depths far beyond those 
typically reported for QAOA. We find that the 
effectiveness of the trained protocol in steering 
the system toward the ground state is rooted in a 
pitchfork bifurcation of the Floquet map, which 
destabilizes the equatorial fixed point and drives 
the spins toward the polar fixed points. 
Finally, we develop a dynamical mean field theory 
in which the problem reduces to an effective 
Landau-Lifshitz dynamics for a single spin 
in a stochastic magnetic field. For optimized 
parameters, the magnetic field covariance matrix 
exhibits a low effective rank, in stark contrast 
to the higher-rank noise structure of the quantum case, 
which explains why the classical dynamics is easier 
to control than its quantum counterpart.

\section{The Variational Iterative Rotation Algorithm}
\label{sec:viral}

\begin{figure}[h!]
	\centering
	\includegraphics[width=\textwidth]{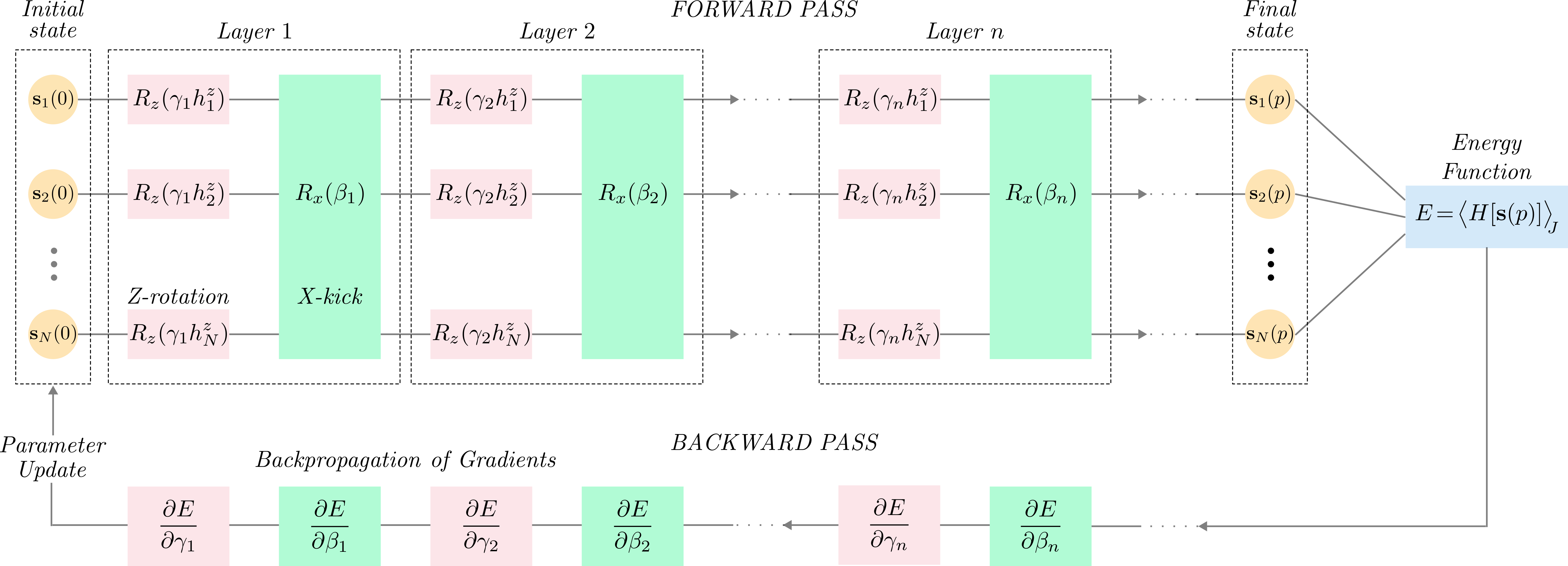}
	\caption{
    \justifying
    {\bf The Variational Iterative Rotation Algorithm.} 
    {\it FORWARD PASS:} The dynamics is initialized in 
    a configuration in which all unit spins $\mathbf{s}_i(0)$ 
    lie in the $xz$-plane at a tilt angle $\alpha$ from 
    the equator, uniformly for the SK model and with random 
    signs of the $z$-component for Max-Cut (see text). 
    Each layer $n$ applies to each spin a site-dependent 
    rotation about the $z$-axis by the angle $\gamma_n h_i^z$, 
    set by the local field $h_i^z$ (pink boxes), followed by a 
    global rotation about the $x$-axis by the angle $\beta_n$  
    (green boxes). After $p$ layers, the average energy 
    $E=\langle H[\mathbf{s}_p]\rangle_J$ is evaluated on 
    the final state. 
    {\it BACKWARD PASS:} Since the dynamics is 
    differentiable, the gradients of $E$ with respect 
    to all the variational parameters 
    $\{\alpha, \gamma_n, \beta_n\}$ are computed exactly 
    by backpropagation from the final state through the 
    layers (see Supplementary Sec.~\ref{sec:backprop}), 
    and the parameters are updated iteratively until 
    convergence. 
}
\label{fig:figure1}
\end{figure}

We begin by introducing the algorithm and its 
implementation. VIRAL is fundamentally a dynamical 
system composed of $N$ classical tops, i.e., 
3-component unit vectors 
\begin{equation}
\mathbf{s}_i=(s_i^x,s_i^y,s_i^z)\ ,\ \ \ {\rm with}\ \ |\mathbf{s}_i|=1\ ,
\end{equation}
representing the orientation of each top evolving 
under a deterministic Hamiltonian dynamics.  
To explain the implementation of the algorithm, 
we focus on the canonical problem of finding the 
ground state of the Sherrington-Kirkpatrick (SK) 
model, defined by the cost function 
\begin{equation}
H[\sigma]= \sum_{i<j} J_{ij}\,\sigma_i\,\sigma_j\ ,
\end{equation}
where $\sigma_i=\pm1$ are discrete Ising variables 
and the couplings $J_{ij}$ are i.i.d. Gaussian random 
variables with zero mean and variance $1/N$. 
Within our framework, the Ising variables are promoted 
to the $z$-components of the classical spins 
$\sigma_i \to s_i^z$, which maps the SK cost function
onto the Hamiltonian of $N$ interacting tops 
\begin{equation}
H[\mathbf{s}] = \sum_{i<j} J_{ij}\,s_i^z\,s_j^z\ .
\label{eq:Htop}
\end{equation}
The ground state of $H[\mathbf{s}]$ coincides 
exactly with the SK ground state when the tops 
are fully polarized along the $z$-axis, 
$s_i^z = \pm 1$. 
Our protocol initializes all $N$ spins identically 
in the $xz$-plane at an angle $\alpha$, such that 
$\mathbf{s}_i(0) = (\cos\alpha,\,0,\,\sin\alpha)$,  
with the angle $\alpha$ serving as the first variational 
parameter. The system then evolves through $p$ discrete 
layers, as illustrated in Fig.~\ref{fig:figure1}. 
Each layer $n$ comprises two successive rotations: 
first, a site-dependent rotation $R_z(\gamma_n)$ 
around the $z$-axis, where each spin $i$ precesses 
by an angle $\gamma_n h_i^z$ determined by its local 
field $h_i^z = \sum_{j \neq i} J_{ij}\, s_{j}^z$; 
second, a uniform rotation $R_x(\beta_n)$ around 
the $x$-axis, kicking all spins by a common angle 
$\beta_n$. After $p$ layers, the final state is 
given by 
\begin{equation}
\mathbf{s}(p) = R_x(\beta_p)\, R_z(\gamma_p) \cdots R_x(\beta_1)\, R_z(\gamma_1)\, \mathbf{s}(0)\ , 
\label{eq:map}
\end{equation}
and the cost function is the disorder-averaged 
energy evaluated on this final configuration 
\begin{equation}
E=\langle H[\mathbf{s}(p)]\rangle_J\ ,    
\end{equation}
which is minimized with respect to the $2p+1$ 
variational parameters 
$\{\alpha,\gamma_1,\beta_1,...,\gamma_p,\beta_p\}$. 
The central practical advantage of this formulation 
is that the layered map in Eq.~\eqref{eq:map} is not 
only deterministic but also differentiable with respect 
to all the variational parameters. The gradients of 
the cost function can therefore be computed exactly by 
backpropagating the derivatives 
$\partial E/\partial \mathbf{s}_i(n)$ through the 
layers, from the final state to the initial one 
(the explicit backward recursion is derived  
in Supplementary Sec.~\ref{sec:backprop}). 
Each training iteration consists of a forward pass 
evolving the spins and a backward pass propagating 
the derivatives, 
whose cost scales as the product of the circuit 
depth and the number of couplings, which is 
$O(pN^2)$ for the SK model. 
This is in stark contrast with QAOA, where each 
gradient component must be estimated from repeated 
measurements on a quantum computer or from classical 
simulations whose cost grows exponentially with the 
system size. Even the specialized methods that evaluate 
the QAOA performance exactly in the infinite-size 
limit~\cite{farhi2022, basso22} incur a cost exponential 
in the circuit depth, limiting them to $p=20$ at most, 
while approximate matrix-product-state simulations 
extend the optimization to $p=80$~\cite{boulebnane2025evidencequantumapproximateoptimization}. 
Our backpropagation scheme, whose cost remains 
linear in $p$, allows us instead to fully 
optimize the protocol at depths up to $p=130$, training 
directly on 24 SK realizations of size $N=2^{14}$ 
spins. 

\begin{figure}[h]
	\centering
	\includegraphics[width=\textwidth]{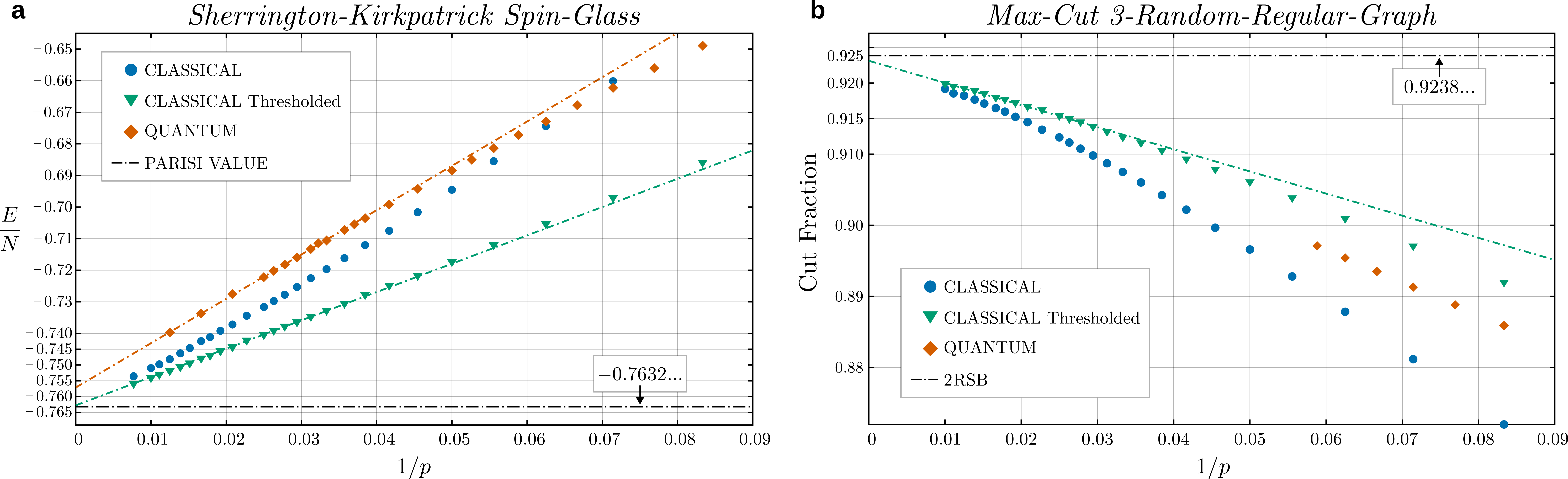}
	\caption{
    \justifying
    {\bf Performance of VIRAL on the SK model and 
    on Max-Cut for random 3-regular graphs.}    
    {\bf(a)}, Energy density $E/N$ reached after 
    $p$ layers of classical (blue circles) or 
    quantum (red diamonds) evolution on the SK 
    model, as a function of $1/p$. The classical 
    results were obtained by optimizing the 
    variational parameters on $24$ instances with 
    $N=2^{14}$ spins, for circuit depths up to $p=130$; 
    the green triangles include the final thresholding 
    step, $\sigma_i=\mathrm{sign}(s_i^z)$.  
    Quantum results are taken from Refs.~\cite{farhi2022,basso22,boulebnane2025evidencequantumapproximateoptimization}. Dashed lines denote linear fits 
    to the $p\geq30$ data, and the dash-dotted 
    horizontal line marks the exact Parisi 
    ground state energy density $-0.76321...$~\cite{crisanti2002analysis}. 
    {\bf(b)}, Cut fraction reached after $p$ layers 
    for Max-Cut on random 3-regular graphs with 
    $N=10^6$ vertices, whose variational parameters were 
    optimized over $10$ instances for depths up to $p=100$ 
    (same color coding as in {\bf(a)}). 
    Quantum results, computed exactly in the infinite-size 
    limit, are taken from Ref.~\cite{farhi2025lower}. 
    The dash-dotted horizontal line marks the 2RSB upper 
    bound $0.9238...$~\cite{harangi2025rsb} on the optimal 
    cut fraction. 
}
\label{fig:figure2}
\end{figure}

The numerical simulations shown in Fig.~\ref{fig:figure2}a 
confirm that the optimized protocol steers the tops 
toward the poles, progressively freezing them into an 
Ising configuration, with an energy density that 
decreases linearly in $1/p$. The performance is further 
improved by a final thresholding step, which rounds each 
spin to the nearest pole, $\sigma_i={\rm sign}(s_i^z)$ 
(green triangles in Fig.~\ref{fig:figure2}a). Linear 
fits to the $p\geq30$ data extrapolate to 
$E/N=-0.7628(6)$, consistent with the exact 
Parisi ground state value of 
$-0.76321...$~\cite{crisanti2002analysis}, 
and notably outperforming QAOA, whose data extrapolate 
to the higher value $E/N =-0.7571(5)$.
For $p=1$, the optimal parameters can be derived 
analytically (see Supplementary Sec.~\ref{sec:singleLayerSI}),  
$\alpha=\beta=\pi/4,\ \gamma=-\sqrt{2}$, 
yielding the variational energy $E/N=-1/(4\sqrt{e})=-0.1516...$, 
exactly half of the QAOA value $E/N=-1/(2\sqrt{e})=-0.3032...$ 
at the same depth (both values lie outside the scale 
of Fig.~\ref{fig:figure2}a, which focuses on the large-$p$ 
limit).

To demonstrate that the performance of VIRAL is not 
tied to fully-connected models, we also benchmark 
it on Max-Cut for random 3-regular graphs, a canonical 
sparse optimization problem for which the QAOA performance 
is known exactly in the infinite-size limit~\cite{farhi2025lower} 
for depths up to $p=17$. 
Given a graph with edge set $\mathcal{E}$, the problem 
is encoded in the Hamiltonian of Eq.~\eqref{eq:Htop} by 
taking $J_{ij}=1$ on the edges of the graph and zero 
otherwise, so that the maximum cut corresponds to the 
ground state, with the cut fraction given by 
\begin{equation}
\text{Cut fraction}=\frac{1}{2} - \frac{E}{2|\mathcal{E}|}\ .     
\end{equation}
In contrast with the SK model, where the random couplings 
render the local fields inhomogeneous from the very first 
layer, in a regular graph with uniform couplings and 
identically initialized spins, the local fields are all 
the same, and stay the same at all times, locking the 
system onto a common trajectory whose cut fraction never 
exceeds 1/2, no better than a random assignment. 
We therefore break this symmetry in the initial condition, 
initializing each spin as 
$\mathbf{s}_i(0)=(\cos\alpha,0,\epsilon_i\sin\alpha)$, 
where the $\epsilon_i=\pm1$ are i.i.d. random signs with equal 
probability, while the angle $\alpha$ remains a variational 
parameter controlling the magnitude of the initial tilt 
from the equator. Each training instance thus consists 
of a random graph together with a realization of the signs 
$\{\epsilon_i\}$, and the single set of optimized 
parameters generalizes across both the graphs and the sign 
realizations. 

Since the couplings are sparse, the cost of a training 
iteration drops to $O(p N)$, which allows us to optimize 
the protocol directly on 10 random graphs with $N=10^6$ 
vertices for depths up to $p=100$, orders of magnitude 
beyond the sizes accessible to quantum simulations. 
As shown in Fig.~\ref{fig:figure2}b, the thresholded 
protocol outperforms the exact infinite-size QAOA 
values~\cite{farhi2025lower} at all available depths, 
with a cut fraction that extrapolates linearly in $1/p$ 
to $0.9231(5)$, just below the 2RSB upper bound 
of $0.9238...$~\cite{harangi2025rsb}. Evaluated directly 
on the depth-$p$ tree, where the comparison with the 
QAOA is exactly defined, VIRAL moreover yields rigorous 
lower bounds on the maximum cut fraction of large-girth 
3-regular graphs that exceed the QAOA bounds at every 
depth $p\geq 4$, raising the best asymptotic lower bound 
from $0.912$ (obtained in Refs.~\cite{csoka2015invariant, gamarnik2018max, farhi2025lower}) to $0.9127$~\cite{morone2026whither}.

\section{Equivalence to the dynamics of classical kicked tops}
\label{sec:kickedtops}

The layered protocol described above is equivalent 
to the dynamics of $N$ coupled classical tops 
governed by the time-dependent Hamiltonian
\begin{equation}
H(t) = \gamma(t)\, \sum_{i<j=1}^N J_{ij}\, s_i^z\, s_j^z 
    + \beta(t) \sum_{i=1}^{N} s_i^x\ ,
\label{eq:manybodyH}
\end{equation}
where the interaction strength $\gamma(t)$ is a 
piecewise-constant function of time, $\gamma(t) = \sum_{n=1}^{p} \gamma_n\,{\bf 1}{[n-1 \leq t < n]}$, modulating 
the $zz$-coupling at each layer, and the 
transverse field $\beta(t)$ acts as a train of 
delta-function kicks, $\beta(t) = \sum_{n=1}^{p} \beta_n\,\delta(t - n)$, where $\beta_n$ is the kick strength at layer $n$. 
These two terms play complementary roles: the 
$zz$-interaction drives a site-dependent precession 
{\it around} the $z$-axis, while the transverse kick rotates 
the resulting $y$-component {\it toward} the $z$-axis, 
thus updating the spin polarizations. 
This continuous Hamiltonian generates precisely 
the discrete map in Eq.~\eqref{eq:map} at each layer, 
$\mathbf{s}_i(n+1) = R_x(\beta_n)\, R_z(\gamma_n h_{i,z}^{(n)})\, \mathbf{s}_i(n)$, where $h_{i,z}^{(n)} = \sum_{j \neq i} J_{ij} s_{j}^z(n)$ is the local field acting on the $z$-component 
of spin $i$. 
To understand how the algorithm steers the spins from 
their initial state to the SK ground state, we analyze 
the fixed-point structure of the map in 
Eq.~\eqref{eq:map}. Two families of fixed points 
are crucial here: the $2^N$ {\it equatorial} fixed 
points where $\mathbf{s}_i^* = (\pm 1, 0, 0)$, which 
exist for any value of $\gamma$ and $\beta$, and the 
$2^N$ {\it polar} (Ising) fixed points where 
$\mathbf{s}_i^* = (0, 0, \pm 1)$, which exist only 
for $\beta=0$. Since the system is initialized near the 
uniform equatorial fixed point $(1,0,0)$, the success 
of the algorithm relies on destabilizing this state 
and steering the system toward the correct polar 
fixed point, i.e. the SK ground state configuration.

\begin{figure}[h!]
	\centering
	\includegraphics[width=\textwidth]{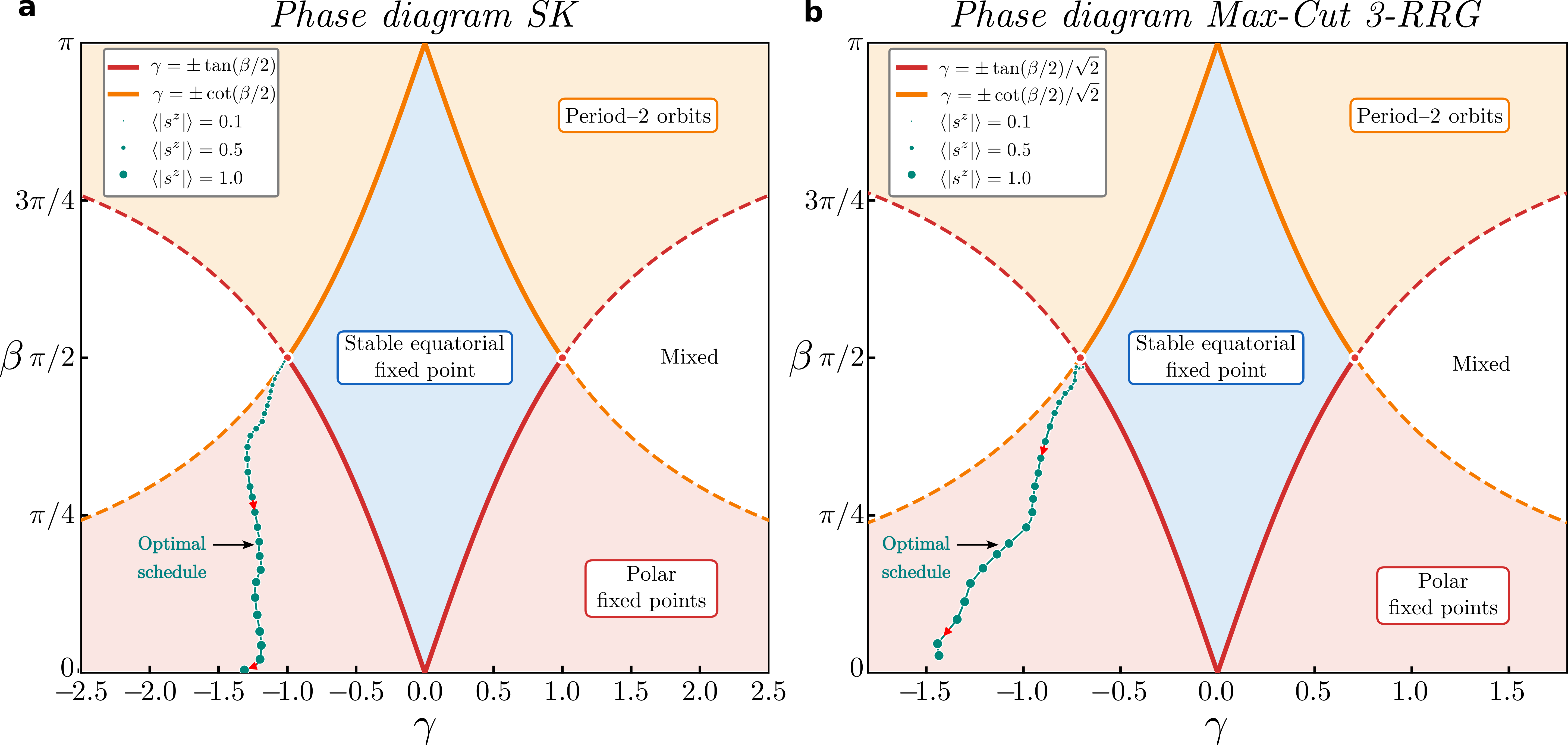}
	\caption{
    \justifying
    \textbf{Stability phase diagram of the 
    VIRAL map and the optimal schedule.}
    {\bf (a)}, Phase diagram in the $(\gamma, \beta)$ 
    plane for the SK model. In the blue region 
    the equatorial fixed point is stable (all 
    normal-mode frequencies $\omega_k$ are real). 
    The red curves mark the pitchfork boundaries 
    $\gamma = \pm\tan(\beta/2)$, beyond which 
    polar fixed points emerge (red region), and 
    the orange curves mark the period-doubling 
    boundaries $\gamma = \pm\cot(\beta/2)$, 
    beyond which period-2 orbits appear (orange region). 
    In the white regions both instabilities can 
    occur. The four boundaries intersect at the two 
    multicritical points 
    $(\beta, \gamma) = (\pi/2, \pm1)$ (red dots). 
    The green circles show the optimal schedule 
    $\{(\gamma_n, \beta_n)\}_{n=1}^{p}$ for 
    $p=130$, which starts near the multicritical
    point and crosses the pitchfork boundary, 
    descending deep into the polar fixed-point 
    region as $\beta\to0$. 
    The size of each circle is proportional to the 
    mean polarization $\langle|s_i^z|\rangle$ at that 
    layer ($N=16384$, $10$ disorder samples, only 
    layers with $\langle|s_i^z|\rangle\geq0.1$ are shown), 
    and the red arrows give the direction of increasing 
    layer index $n$.    
    {\bf (b)}, The same for Max-Cut on random 3-regular 
    graphs, with all boundaries and the multicritical 
    points rescaled by $1/\sqrt{2}$. The optimal schedule 
    is shown for $p=100$ ($N=10^6$, $10$ disorder samples). 
    The full optimized schedules $\{\beta_n,\gamma_n\}$ 
    as functions of the layer index are reported in 
    Supplementary Fig.~\ref{fig:figureS1}. 
    }
    \label{fig:figure3}
\end{figure}

Linearizing the map around the equatorial 
fixed point, we find that in the eigenbasis 
of the coupling matrix $J$ the evolution 
factorizes into $N$ independent modes, 
one for each eigenvalue $\lambda_k$ of $J$ (see 
Supplementary Sec.~\ref{sec:manybodykickedtopSI}), 
with frequencies given by 
$\omega_k = \cos^{-1}\big[\cos\beta + 
(\gamma\lambda_k/2)\sin\beta\big]$. 
Thus, the equatorial fixed point remains stable provided 
all normal mode frequencies are real, yielding the 
stability condition 
\begin{equation}
|2\cos\beta + \gamma\lambda_k\sin\beta| < 2\ \ \ \forall k,\ \ \ 
{\rm Stability\ of\ the\ equatorial\ FP}\ 
\mathbf{s}_i^* = (1,0,0)\ .
\label{eq:stabilityFPequator}
\end{equation}
Since the eigenvalues follow the 
Wigner semicircle law with support 
$\lambda \in [-2, 2]$, the two edges of the 
spectrum dictate the onset of the instability 
by saturating the inequality in 
Eq.~\eqref{eq:stabilityFPequator}. 
Depending on the parameters, two types of instability 
can occur: {\it (i)} a pitchfork bifurcation, in which 
a soft mode frequency vanishes ($\omega\to0$) and new 
stable fixed points with $s_i^z\neq0$ emerge above 
and below the equator; {\it (ii)} a period-doubling 
instability, in which a normal mode frequency 
reaches $\omega\to\pi$, so that the mode amplitude 
changes sign at every kick, creating a period-2 orbit. 
The stability boundaries are defined by 
$\gamma=\pm\tan(\beta/2)$ for the pitchfork bifurcations, 
and by $\gamma=\pm\cot(\beta/2)$ for the period 
doubling instabilities, as shown in the phase 
diagram in Fig.~\ref{fig:figure3}a.

The optimal trajectory followed by our algorithm, 
shown in Fig.~\ref{fig:figure3}a, starts near the 
multicritical point $(\beta, \gamma)=(\pi/2,-1)$, 
where the normal-mode frequencies span the entire 
range $\omega\in[0,\pi]$, and crosses the pitchfork 
line into the polar fixed point region as $\beta$ 
is reduced. In the early layers, the transverse 
kick is large, $\beta_n\sim\pi/2$, and acts as a 
violent mixer. For spins near the equator, the 
$z$-rotations generate $y$-components that the 
$x$-kicks convert into $z$-components. However, 
the kicks are so large that they prevent any 
$z$-polarization from building up, since whatever 
polarization is generated in one layer is immediately 
rotated back toward the equator by the next kick. 
This aggressive phase is needed to efficiently 
explore the phase space and to prevent a premature 
freezing of the spins' $z$-component into sub-optimal 
configurations. Then, as $\beta$ decreases across layers, 
the mixing effect weakens and the $z$-polarization 
is increasingly preserved from one layer to the next, 
allowing the spins to progressively settle in an 
Ising-like configuration. 

The same analysis extends directly to Max-Cut, since 
the stability condition in Eq.~\eqref{eq:stabilityFPequator} 
depends on the coupling matrix only through its spectrum. 
For a random 3-regular graph, the spectrum of the adjacency 
matrix consists of a bulk that follows the Kesten-McKay 
law with support $\lambda\in[-2\sqrt{2},2\sqrt{2}]$, 
together with the isolated Perron-Frobenius eigenvalue 
$\lambda=3$ associated with the uniform eigenvector. 
The random-sign initialization of Sec.~\ref{sec:viral}, 
$s_i^z(0)=\epsilon_i\sin\alpha$, has a vanishing overlap 
with the uniform eigenvector, of order $N^{-1/2}$, so 
the Perron mode is effectively never seeded and the stability 
of the equatorial fixed point is governed by the edges of 
the bulk spectrum alone. 
The stability boundaries are then those of the SK case 
rescaled by $1/\sqrt{2}$, namely $\gamma=\pm\tan(\beta/2)/\sqrt{2}$ 
for the pitchfork bifurcation and $\gamma=\pm\cot(\beta/2)/\sqrt{2}$ 
for the period-doubling instability, with the multicritical 
points moving to $(\beta,\gamma)=(\pi/2,\pm1/\sqrt{2})$, 
as shown in Fig.~\ref{fig:figure3}b. 
The optimal schedule, also shown in Fig.~\ref{fig:figure3}b, 
follows the same strategy as in the SK case. The softening 
mode now sits at the lower edge of the Kesten-McKay spectrum, 
$\lambda\to-2\sqrt{2}$. This instability tends to drive 
neighboring spins toward opposite poles, and since anti-aligned 
neighbors are precisely the cut edges, the polar fixed point 
reached by the optimized dynamics encodes a near-maximal cut.

\subsection{Effective dynamics}


We now return to the SK model, whose infinite-range Hamiltonian allows the many-body problem in 
Eq.~\eqref{eq:manybodyH} to be exactly reduced 
to the effective dynamics of a single spin 
in a magnetic field with a fluctuating $z$-component~\cite{sompolinsky1982relaxational} 
(see Supplementary Sec.~\ref{sec:viralDMFT}). 
The evolution is described by the continuous 
Landau-Lifshitz equation
\begin{equation}
\frac{d\mathbf{s}}{dt}= \mathbf{g}(t)\times \mathbf{s}\ ,
\label{eq:effectiveLLG}
\end{equation}
where the transverse component of the effective 
field is the original train of delta-function 
kicks, $g^x(t)=\beta(t)$, the $y$-component 
vanishes identically, $g^y(t)=0$, and the 
$z$-component $g^z(t)$ is a Gaussian stochastic 
process with zero mean and covariance 
\begin{equation}
\langle\langle g^z(t)g^z(t')\rangle\rangle = \gamma(t)\gamma(t')\langle\langle s^z(t)s^z(t')\rangle\rangle=\gamma(t)\gamma(t')\Sigma(t,t')\ ,
\label{eq:noise_def}
\end{equation}
where the double brackets 
$\langle\langle\cdots\rangle\rangle$ denote the 
average over the realizations of $g^z(t)$, 
replacing the disorder average of the original 
many-body problem. Equation~\eqref{eq:noise_def} 
is thus a self-consistency condition, since 
the covariance of $g^z(t)$ is fixed by the 
correlations of the very trajectory $s^z(t)$ 
it generates. 

Although Eq.~\eqref{eq:effectiveLLG} describes 
a continuous-time process, the discrete nature 
of the transverse kicks greatly simplifies the 
effective dynamics. Since $g^x(t)=0$ between 
consecutive kicks, the $z$-component of the spin 
is constant in each interval $t\in(n-1,n)$, and 
the accumulated effect of the stochastic field 
$g^z(t)$ during the $n$-th layer reduces to a 
single random rotation angle  
\begin{equation}
\theta_n = \int_{n-1}^n g^z(t)\ dt\ .
\end{equation}
The angles $\theta_n$ are Gaussian random variables 
with zero mean and covariance 
\begin{equation}
\langle\langle \theta_m \theta_n\rangle\rangle = 
\gamma_m\gamma_n \langle\langle s^z(m-1)s^z(n-1)\rangle\rangle = 
\gamma_m\gamma_n\Sigma(m-1,n-1)\ ,
\label{eq:noise_covarariance}
\end{equation}
where $s^z(n-1)$ denotes the $z$-component of 
the spin right after the $n-1$-th kick. Crucially, 
the covariance of $\theta_n$ involves only spin 
states from the layers preceding $n$. The effective 
dynamics can therefore be sampled exactly, layer by 
layer, as a sequence of rotations with each angle 
drawn conditionally on the previous ones (see 
Supplementary Sec.~\ref{sec:effective-single-spin-LLG}), 
bypassing the solution of the self-consistency condition 
Eq.~\eqref{eq:noise_def} in continuous time. 

Once the dynamics is reduced to a discrete sequence of 
rotations driven by the stochastic angles $\theta_n$, 
the goal of the algorithm is to tune the control parameters 
$\gamma_n$ and $\beta_n$ to minimize the expected energy. 
In the thermodynamic limit, the disorder-averaged energy 
can be evaluated exactly at the final layer $p$. 
The resulting energy density, derived in 
Supplementary Sec.~\ref{sec:pathIntegralSI}, is given by
\begin{equation}
\lim_{N\to\infty}\frac{E}{N} = \sum_{n=1}^p \gamma_n\, C(n)\, R(n)\ ,
\label{eq:energydecomposition}
\end{equation}
where $C(n)\equiv \langle\langle s^z(n-1)s^z(p) \rangle\rangle$ 
is the temporal correlation function between the final 
state and the state in layer $n$, and 
$R(n)\equiv \langle\langle \partial s^z(p)/\partial\theta_n \rangle\rangle$ is the response function of the final 
state to a perturbation of the angle $\theta_n$ in layer $n$. 

\begin{figure}[h!]
	\centering
	\includegraphics[width=\textwidth]{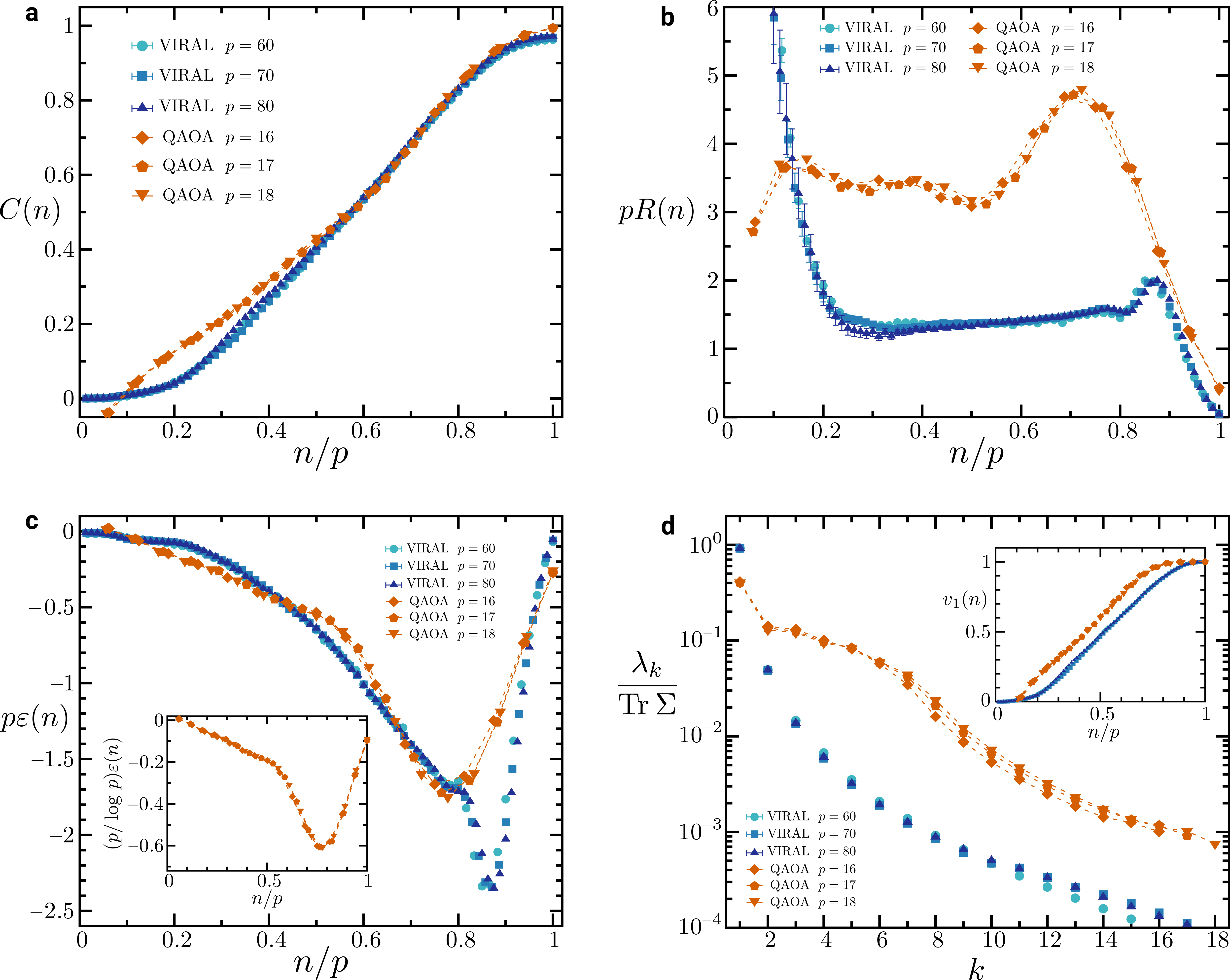}
	\caption{
    \justifying
    {\bf Dynamical mean-field analysis of the SK 
    model: correlation, response, layer energy, 
    and noise spectrum of the classical dynamics 
    compared with QAOA.} 
    All data refer to the SK model and are plotted 
    vs the rescaled layer index $n /p$. The VIRAL 
    data are obtained numerically on system sizes 
    of $N=4096$ spins averaged over $64,000$ disorder 
    samples (error bars are s.e.m. and are smaller 
    than symbols where not visible), using the 
    optimized schedules at $p=60$, $p=70$, $p=80$. 
    The QAOA data are exact infinite-size values at 
    $p=16$, $p=17$, $p=18$, obtained by iterating the 
    saddle point equations  of Farhi {\it et al.}~\cite{farhi2022} 
    with the optimized angles of Basso {\it et al.}~\cite{basso22} 
    (see Supplementary Sec.~\ref{sec:qaoaG}).
    {\bf (a),} Correlation $C(n)=\langle\langle s^z(p)s^z(n-1)\rangle\rangle$ between the final 
    $z$-component of the spin and the one on which 
    layer $n$ acts. For QAOA, $C(n)=\mathrm{Re}(G_{0n})$ 
    is the symmetrized correlation between the measured 
    output $Z$ and the $Z$ at layer $n$. 
    {\bf (b),} Response $R(n)=\langle\langle\partial s^z(p)/\partial\theta_n\rangle\rangle$ of the final 
    $z$-component to the rotation angle $\theta_n$ 
    of layer $n$. For QAOA, $R(n)=-2\mathrm{Im}(G_{0n})$ 
    is the corresponding response. Both are multiplied 
    by $p$, so that different depths can be compared 
    (note the collapse  of the classical data). 
    {\bf (c),} Layer energy $\varepsilon(n)=\gamma_n C(n)R(n)$ 
    of Eq.~\eqref{eq:energydecomposition}, multiplied 
    by $p$. The sum over $n$ of $\varepsilon(n)$ is the 
    classical energy density $E/N$ and the QAOA $-\nu_p$. 
    Inset: the QAOA curves rescaled by $p/\log p$ instead 
    of $p$, as suggested in Ref.~\cite{sels2026absence}, 
    seem to collapse better. 
    {\bf (d),} Eigenvalues $\lambda_k$ of the covariance 
    $\Sigma(t,t')=\langle\langle s^z(t)s^z(t')\rangle\rangle$ 
    with $t,t'=0,\dots,p-1$, whose product with 
    $\gamma(t)\gamma(t')$ is the noise covariance 
    of Eq.~\eqref{eq:noise_covarariance}, normalized 
    by the trace. For QAOA, $\Sigma=\mathrm{Re}(G_{tt'})$ 
    on the forward branch of the contour, $t,t'=1,\dots,p$. 
    The leading eigenvalue carries $92\%$ of the trace 
    for VIRAL and $40\%$ for QAOA. Inset: leading eigenvector 
    $v_1(n)$ of $\Sigma$, normalized to its maximum. For 
    VIRAL, $v_1(n)$ coincides with $C(n)$ of panel ${\bf (a)}$ 
    to within $5\%$, as it should for a covariance of 
    rank one.  
    }
    \label{fig:figure4}
\end{figure}

Equation~\eqref{eq:energydecomposition} reveals the core 
of the optimization scheme. The total energy is the sum 
of contributions from each layer $n$, each given by the 
product of three factors: the interaction strength $\gamma_n$, 
the correlation function $C(n)$, and the response function 
$R(n)$ (the latter two are the central objects of the mean-field 
dynamics of glassy systems~\cite{sompolinsky1982relaxational, cugliandolo1993analytical}). 
Remarkably, the same decomposition holds for the QAOA (see 
SupplementarySec.~\ref{sec:qaoaG}), where the correlation 
$C(n)$ is the symmetrized correlator of the operators 
$Z(n-1)$ and $Z(p)$, and the response $R(n)$ is 
their commutator, as summarized in Table~\ref{tab:viral-qaoa}. 
This correspondence puts the two protocols on exactly the 
same footing, thus enabling the direct quantitative 
comparison shown in Fig.~\ref{fig:figure4}.

Correlation and response operate in opposition throughout 
the protocol. In the early layers the system exhibits a 
large response, since the spins are near the equator and 
experience large transverse kicks, so that a small perturbation 
of the angle $\theta_n$ drastically alters the subsequent 
trajectory. 
The correlation is instead negligible, because these same 
large kicks effectively erase the memory of the spin, and 
the early state $s^z(n-1)$ carries little information about 
its final destination $s^z(p)$. 
In the late layers the roles are reversed. As $\beta_n$ 
decreases, the $z$-polarization is increasingly preserved 
and the spins commit to their Ising positions, becoming 
insensitive to further rotations, so the response vanishes, 
while the current orientation becomes perfectly correlated 
with the final state, pushing $C(n)$ toward $1$. The optimal 
schedule balances the loss of susceptibility with the gain 
in memory, and the energy is built up across all layers, 
with the largest contributions $\varepsilon(n)$ coming 
from the intermediate layers where the product of correlation 
and response is largest (Fig.~\ref{fig:figure4}c). 
The comparison with QAOA sharpens this picture. The 
correlation profiles of the two protocols are nearly 
indistinguishable when plotted against the rescaled 
layer index $n/p$ (Fig.~\ref{fig:figure4}a), and the 
two dynamics build up their energy with the same 
profile (Fig.~\ref{fig:figure4}c). The crucial difference 
lies in the response, in that the classical response 
collapses when multiplied by $p$, i.e. $R(n)\sim 1/p$, 
whereas the quantum response decays more slowly with 
depth (Fig.~\ref{fig:figure4}b), consistent with the 
logarithmically slower convergence of the QAOA 
energy derived in Ref.~\cite{sels2026absence} and 
with the extrapolation of Fig.~\ref{fig:figure2}a.

Within the dynamical mean field theory, the 
controllability of the protocol is encoded in 
the spectrum of the noise covariance $\Sigma(t,t')$, 
shown in Fig.~\ref{fig:figure4}d. In the classical 
case the spectrum is dominated by a single eigenvalue, 
carrying $92\%$ of the trace, whose eigenvector 
coincides with the correlation profile $C(n)$ (inset 
of Fig.~\ref{fig:figure4}d), as expected for a matrix 
of rank one. The near rank-one structure, combined 
with the self-consistency condition Eq.~\eqref{eq:noise_def}, 
implies that the stochastic field reduces to 
$g^z(t)\approx s^z(t)\xi$, with a single random variable 
$\xi$ shared by all layers. As such, the dynamics 
is equivalent to a purely Hamiltonian evolution with 
effective Hamiltonian 
\begin{equation}
    H_{\rm eff}=\frac{\xi \gamma(t)}{2} (s^z)^2+\beta(t)s^x\ ,
\end{equation}
whose Floquet map possesses a family of stable fixed 
points, parametrized by $\beta$, that continuously connects 
the equatorial fixed point to the poles through the same 
pitchfork bifurcation found in the full many-body analysis. 
The efficiency of the control is thus tied to the existence 
of a smooth trajectory of stable fixed points, which at 
large $p$ can be followed adiabatically by slowly reducing 
$\beta$ from $\pi/2$ to $0$ at nearly fixed $\gamma$, precisely as 
realized by the optimal schedule shown in Fig.~\ref{fig:figure3}a. 
The quantum case is fundamentally different. The counterpart 
of $\Sigma$ is the symmetrized correlator $\mathrm{Re}\big[G(t,t')\big]$ 
(see Table~\ref{tab:viral-qaoa}), whose diagonal is 
pinned to $\mathrm{Re}\big[G(t,t)\big]=1$ at all times 
by the identity $Z_i^2=1$, so that quantum fluctuations 
are of order one throughout the evolution. In contrast, 
the classical variance $\Sigma(t,t)$ grows from the small 
initial tilt $\sin^2\alpha$ toward $1$ only as the spins 
polarize. For $\mathrm{Re}\,G$ to be close to rank one, 
all of its off-diagonal entries would also have to be 
close to one, which would require $Z$ to be conserved 
during the entire evolution. But a conserved $Z$ would 
freeze the energy at its initial value, so any protocol 
that lowers it must not conserve $Z$. The quantum covariance 
therefore necessarily retains several eigenvalues of 
order one, as confirmed in Fig.~\ref{fig:figure4}d, 
where the leading eigenvalue carries only $40\%$ of the 
trace. In other words, the classical spin feels an 
initially weak field with a single frozen random component, 
which a deterministic schedule can steer around, while 
the qubit feels a strong field built from many independent 
random components, so that a single schedule fixed in 
advance cannot steer all realizations of the dynamics 
efficiently. 
At short depths this noise is in fact beneficial, and QAOA 
outperforms the unthresholded classical protocol at small 
$p$. Conversely, at large depths the noise hampers the 
control and produces the markedly slower convergence of 
QAOA observed in Fig.~\ref{fig:figure2}a.

\begin{table*}[t]
\centering
\renewcommand{\arraystretch}{1.8}
\setlength{\tabcolsep}{6pt}
\begin{tabular}{|l|m{0.33\textwidth}|m{0.33\textwidth}|}
\hline
& VIRAL (classical) & QAOA (quantum) \\
\hline
Layer $n$ & $\mathbf{s}_i(n)=R_x(\beta_n)R_z(\gamma_n h_i^z)\mathbf{s}_i(n-1)$, 
\vspace{2mm} 
\newline
$h_i^z = \sum_j J_{ij} s_j^z$
& $|\psi(n)\rangle = U_n |\psi(n-1)\rangle$, 
\vspace{2mm}
\newline 
$U_n = e^{-i\beta_n \sum_i X_i} e^{-i\gamma_n \sum_{i<j} J_{ij} Z_iZ_j}$
\\
\hline
Spin $i$ after $t$ layers & $s^z_i(t)$ & $Z_i(t) = W_t^\dagger Z_i W_t$\ ,\ \ \ 
$W_t = U_t\cdots U_1$\\
\hline
Two-time correlator & $\Sigma(t,t') = \langle\langle s^z(t) s^z(t')\rangle\rangle$ & $G(t,t') = \langle Z(t)Z(t')\rangle\ ,\ \ \  t\leq t'$\\
\hline
Correlation $C(n)$ & $\Sigma(n-1,p)=\langle\langle s^z(n-1) s^z(p)\rangle\rangle$ & $\mathrm{Re}\big[ G(n-1,p)\big] = 
\frac{1}{2}\langle\{Z(n-1),Z(p)\}\rangle$\\
\hline
Response  $R(n)$   & $\langle\langle \partial s^z(p)/\partial \theta_n\rangle\rangle$ & $-2\,\mathrm{Im}\big[G(n-1,p)\big] = 
i\langle [Z(n-1), Z(p)]\rangle
$\\
\hline
Layer energy $\varepsilon(n)$ & $\gamma_n C(n)R(n)$ & $\gamma_n C(n)R(n)$\\
\hline 
Energy density $E/N$ & $\sum_{n=1}^p\varepsilon(n)$ & $\sum_{n=1}^p\varepsilon(n)$\\
\hline
Covariance (Fig.~\ref{fig:figure4}d) & $\Sigma(t,t')\ ,\ \ \ t,t'=0,...,p-1$ 
& $\mathrm{Re}\big[ G(t,t')\big] = \frac{1}{2}\langle\{Z(t),Z(t')\}\rangle$ \\
\hline 
Stochastic field on one spin & $\langle\langle g^z(t) g^z(t')\rangle\rangle = \gamma(t)\gamma(t')\Sigma(t,t')$ & $\langle\langle g^z(t) g^z(t')\rangle\rangle = \gamma(t)\gamma(t')G(t,t')$\\
\hline
\end{tabular}
\caption{
\justifying
{\bf{Correspondence between the classical and the 
quantum quantities of Fig.~\ref{fig:figure4} for 
the SK model.}}
The variables $s_i^z(t)$ and $Z_i(t)$ refer to spin 
$i$ after $t$ layers ($t=0,...,p$), so that layer 
$n$ acts on $s_i^z(n-1)$ and $Z_i(n-1)$. 
The double brackets $\langle\langle\cdot\rangle\rangle$ 
denote the average over the realizations of the 
classical stochastic field, and $\langle\cdot\rangle$ 
the quantum expectation of the evolved operators 
in the state $|+\rangle^{\otimes N}$, averaged over 
the couplings for $N\to\infty$. 
Note that QAOA rotates the state by $2\gamma_n h_i^z$ 
about $z$ and by $2\beta_n$ about $x$, twice the VIRAL 
angles. 
} 
\label{tab:viral-qaoa}
\end{table*}

To conclude this section we propose nanometer-scale 
magnetic tunnel junction (MTJ) free layers as a 
natural physical platform for VIRAL. At these length 
scales, the free-layer magnetization behaves as a 
single macrospin governed by the classical 
Landau-Lifshitz–Gilbert equation, and, unlike in 
conventional Ising machines with binary degrees of 
freedom, it constitutes a continuous three-dimensional 
vector spin. By engineering the perpendicular magnetic 
anisotropy to compensate the shape anisotropy~\cite{Ikeda2010,Worledge2011,Shaw2015}, the 
macrospin can explore the full Bloch sphere rather than 
being confined to two logical poles~\cite{Kurebayashi2026}. 
The rotations required by the algorithm can be implemented 
with spin-transfer and spin-orbit torques, and applied 
magnetic fields, with pulse trains synchronized by FPGA-based 
electronics, as recently demonstrated experimentally~\cite{Sidi2025,Criss2026}, 
while the spin state can be read out electrically via the 
tunnel magnetoresistance~\cite{duffee2025integrated}. 
The coherent dynamics unfolds on the tens-of-nanoseconds 
timescale set by the intrinsic magnetic damping, which is 
well separated from slower stochastic effects, thus making 
MTJ macrospin arrays a realistic and experimentally grounded 
platform for the layered rotation protocol of VIRAL.

\section{Conclusion}

In this work we have formulated and trained VIRAL, 
a purely classical and deterministic system 
of kicked tops whose layered rotation structure 
mirrors that of the QAOA, and whose control 
parameters are learned by backpropagation through 
the dynamics. The trained protocol outperforms 
its quantum counterpart on both the Sherrington-Kirkpatrick 
model, where the energy density converges to the 
Parisi value linearly in $1/p$, and on Max-Cut for random 
3-regular graphs, where the cut fraction approaches 
the 2RSB upper bound. Evaluated on the tree, VIRAL 
yields rigorous lower bounds on the maximum cut that 
surpass those of the QAOA~\cite{morone2026whither}. 
The fixed-point analysis of the Floquet map shows 
that the optimization is driven by a pitchfork 
bifurcation, with a soft mode that vanishes like 
$\sqrt{\gamma_c-\gamma}$, through which the spins 
are adiabatically transported from the equator to 
the polarized Ising states. 
The dynamical mean field theory puts the classical 
and quantum protocols on the same footing, with 
the energy built layer by layer from the product 
of a correlation and a response function in both 
cases. The comparison identifies the origin of 
the classical advantage: the classical noise 
covariance is nearly rank-one and the effective 
dynamics remains controllable, whereas quantum 
fluctuations, pinned to order one by the operator 
constraint $Z_i^2=1$, necessarily raise the rank 
of the noise and hamper the control. Our results 
thus indicate that the performance of QAOA stems 
from its iterative rotation structure, while the 
quantum fluctuations at its core hinder, rather 
than aid, the optimization. 

As a final remark, we stress that the delta function 
train for $\beta(t)$ is a rudimentary open loop control 
strategy. Owing to the classical nature of the problem, 
it would be straightforward to implement general feedback 
control with little overhead. In particular, in the MTJ 
platform, the magnetization can be read out via the tunnel 
magnetoresistance at GHz rates with minimal perturbation 
to the system, while FPGAs can implement the feedback in 
real time. For problems more complex than the SK model, 
such as $p$-spin models, one can further add controlled 
noise to improve the exploration of the phase space. 
With all of this in mind, one really wonders, where is 
quantum optimization going to present a clear advantage?

\bigskip

\noindent
{\bf \large Data availability}
The data supporting the findings of this 
study can be reproduced with the publicly 
available code (see Code availability) and 
are also available from the corresponding 
author upon request.

\medskip

\noindent
{\bf \large Code availability}
The source code to perform the optimization 
of the variational parameters can be found at 
\url{https://github.com/flavianoM/VARIATIONAL-ITERATIVE-ROTATION-ALGORITHM}. 

\medskip

\noindent
{\bf \large Acknowledgments} 
The Flatiron Institute is a division of the 
Simons Foundation. D.S. was supported by AFOSR 
under Award No. FA9550-21-1-0236 and D.S. and 
A.D.K. were supported by ONR under Award No. 
N00014-23-1-2771. We thank Jonathan Z. Sun for 
helpful discussions of this work.

\medskip


\noindent
{\bf \large Additional information}
Supplementary Materials accompany this paper. 

\medskip

\noindent
{\bf \large Competing interests} 
All authors declare no competing interests. 

\medskip

\noindent
{\bf \large Correspondence} 
should be addressed to F.M. at: fm2452@nyu.edu

\bibliography{ref_general}

\newpage

\onecolumngrid

{\centering

{\normalsize \bf Supplementary Material: Variational Iterative Rotation Algorithm: Combinatorial Optimization with Classical Kicked Tops}

\vspace{0.5cm}

Flaviano Morone$^{1}$, Andrew D. Kent$^{1}$, and Dries Sels$^{2,3}$

\vspace{0.1in}

{\it
$^1$Center for Quantum Phenomena, Department of Physics, New York University, New York, New York 10003, USA\\
$^2$Department of Physics, Boston University,\\ 590 Commonwealth Ave., Boston, Massachusetts 02215, USA\\
$^3$Center for Computational Quantum Physics, Flatiron Institute,\\
162 Fifth Avenue, New York, New York 10010 USA\\
}}

\vspace{1cm}

\setcounter{section}{0}
\renewcommand{\thesection}{S\arabic{section}}
\setcounter{equation}{0}
\renewcommand{\theequation}{S\arabic{equation}}
\setcounter{figure}{0}
\renewcommand{\thefigure}{S\arabic{figure}}
\setcounter{table}{0}
\renewcommand{\thetable}{S\arabic{table}}

\section{Single layer optimization \texorpdfstring{$p=1$}{p=1}}
\label{sec:singleLayerSI}

In this section we derive analytically the 
optimal parameters for the single layer case 
corresponding to $p=1$. 
The initial state corresponds to all spins 
aligned in the same direction in the $xz$ 
plane at an angle $\alpha$ relative to the 
$x$ axis such that 
\begin{equation}
\mathbf{s}_i(0) = 
\begin{pmatrix}
\cos\alpha\\
0\\
\sin\alpha
\end{pmatrix}
\ .
\end{equation}
To generate $\mathbf{s}_i(1)$ we apply 
two rotations to each spin $\mathbf{s}_i(0)$: 
one of angle $\gamma h_i^z$ around the $z$-axis, 
$R_z(\gamma h_i^z)$, and a second one of angle 
$\beta$ around the $x$-axis, $R_x(\beta)$, 
thus obtaining the total rotation $R_{tot}$
\begin{equation}
R_{tot} = \begin{bmatrix}
\cos(\gamma h_i^z) & -\sin(\gamma h_i^z) & 0 \\
\cos(\beta) \sin(\gamma h_i^z) & \cos(\beta) \cos(\gamma h_i^z) & -\sin(\beta) \\
\sin(\beta) \sin(\gamma h_i^z) & \sin(\beta) \cos(\gamma h_i^z) & \cos(\beta)
\end{bmatrix}
\ ,
\label{eq:Rtot}
\end{equation}
where $h_i^z$ is the $z$ component of the 
local field
\begin{equation}
h_i^z = \sum_{k\neq i}J_{ik}s_{k}^z(0) = 
\sum_{k\neq i}J_{ik}\sin\alpha .
\end{equation}
Applying $R_{tot}$ to $\mathbf{s}_i(0)$ we 
find $\mathbf{s}_i(1)$, whose $z$ component is 
given by 
\begin{equation}
s_i^z(1) =  \sin(\alpha)\cos(\beta) + \cos(\alpha)\sin(\beta)\sin(\gamma h_i^z)\ .
\end{equation}
Having found the new state we need to 
evaluate the Energy $H[\mathbf{s}(1)] = 
\sum_{i<j}J_{ij}s_i^z(1)s_{j}^z(1)$ 
and average over the $J_{ij}$. We take the 
average because we want to find the optimal 
variational parameters that are generic, i.e. 
independent of the particular realization of 
the couplings. 
Due to the symmetry of the distribution $P(J_{ij})$ 
the only piece that survives in the average is 
\begin{equation}
\langle H[\mathbf{s}(1)]\rangle_J = 
2\sin(\alpha)\cos(\beta)\cos(\alpha)\sin(\beta)
\sum_{i<j}\langle J_{ij}\sin(\gamma h_i^z)\rangle_J\ .
\end{equation}
Computing the expectation we find 
\begin{equation}
\begin{aligned}
\langle J_{ij}\sin(\gamma h_i^z)\rangle_J &= 
\mathbb{E}_J\Big[J_{ij}\ {\rm Im}\Big(e^{i\gamma h_i^z}\Big)\Big] = \mathbb{E}_J\Big[J_{ij}\ {\rm Im}\Big(e^{i\gamma 
\sum_{k\neq i}J_{ik}\sin(\alpha)}\Big)\Big] = \\
&= \mathbb{E}_J\Big[J_{ij}\ {\rm Im}
\Big(e^{i\gamma J_{ij}\sin(\alpha)}e^{i\gamma
\sum_{k\neq i,j}J_{ik}\sin(\alpha)}\Big)\Big] = \\
&=
\mathbb{E}_J\Big[J_{ij}\ {\rm Im}
\Big(e^{i\tilde{\gamma} J_{ij}}e^{i\tilde{\gamma}
\sum_{k\neq i,j}J_{ik}}\Big)\Big] = 
\mathbb{E}_J\Big[ {\rm Im}
\Big(J_{ij}e^{i\tilde{\gamma} J_{ij}}e^{i\tilde{\gamma}
\sum_{k\neq i,j}J_{ik}}\Big)\Big] = \\
&= \mathbb{E}_J\Big[ {\rm Im}
\Big(-i\frac{d}{d\tilde{\gamma}}e^{i\tilde{\gamma} J_{ij}}\Big)\Big(e^{i\tilde{\gamma}
\sum_{k\neq i,j}J_{ik}}\Big)\Big] = 
-\Big[\frac{d}{d\tilde{\gamma}}\mathbb{E}_J\Big(
e^{i\tilde{\gamma} J_{ij}}\Big)\Big]
\mathbb{E}_J\Big(e^{i\tilde{\gamma}
\sum_{k\neq i,j}J_{ik}}\Big) = \\
&= -\Big(\frac{d}{d\tilde{\gamma}} e^{-\tilde{\gamma}^2/2N}
\Big)e^{-\tilde{\gamma}^2 (N-2)/2N} = 
\frac{\tilde{\gamma}}{N}e^{-\tilde{\gamma}^2 (N-1)/2N}
\end{aligned}
\ ,
\end{equation}
where we defined $\tilde{\gamma}=\gamma\sin(\alpha)$. 
Summing over $i<j$ yields 
\begin{equation}
\sum_{i<j}\langle J_{ij}\sin(\gamma h_i^z)\rangle_J = 
\sum_{i<j}\frac{\tilde{\gamma}}{N}e^{-\tilde{\gamma}^2 (N-1)/2N} = 
\frac{\tilde{\gamma} (N-1)}{2}e^{-\tilde{\gamma}^2 (N-1)/2N}\sim 
\frac{\tilde{\gamma} N}{2}e^{-\tilde{\gamma}^2 /2}\ .
\end{equation}
Thus we obtain in the thermodynamic limit 
\begin{equation}
\lim_{N\to\infty}\frac{\langle H[\mathbf{s}(1)]\rangle_J}{N} = \sin(\alpha)\cos(\beta)\cos(\alpha)\sin(\beta)
\tilde{\gamma} e^{-\tilde{\gamma}^2 /2}\ ,
\label{eq:cost1layer}
\end{equation}
from which we find the optimal variational 
parameters
\begin{equation}
\begin{aligned}
\alpha_{*} = \pi/4\ ,\ \ \ \beta_{*} = \pi/4\ ,\ \ \ \gamma_{*} = -\sqrt{2}\ ,
\end{aligned}    
\end{equation}
and the optimal energy 
\begin{equation}
\lim_{N\to\infty}\frac{\langle H[\mathbf{s}(1)]\rangle_J}{N}\Bigg|_{\alpha_{*}, \beta_{*}, \gamma_{*}} = 
-\frac{1}{4\sqrt{e}} = -0.15163...
\end{equation}

\section{Training viral with backpropagation}
\label{sec:backprop}

The VIRAL layered structure is naturally 
suited to be trained with backpropagation. 
In fact, the spins in final state $\mathbf{s}_i(p)$ 
are differentiable functions of the variational 
parameters, so the gradients of the final 
energy can be computed analytically at each layer. 
To make the notation lighter we write the map 
in the following form 
\begin{equation}
\begin{aligned}
s_i^a(n+1) &= X_n^{ab}\, t_i^b(n)\ ,\\
t_i^b(n) &= Z_{n,i}^{bc}\, s_i^c(n)\ ,
\end{aligned}
\end{equation}
where the sum over repeated upper indices 
is implied and we defined 
\begin{equation}
\begin{aligned}
X_n &\equiv R_x(\beta_n)\ ,\\
Z_{i,n} &\equiv R_z(\theta_{n,i})\ ,\\
\theta_{n,i} &\equiv \gamma_n h_i^{(n)}
\end{aligned}
\end{equation}
The objective of backpropagation is to 
compute 
\begin{equation}
\frac{\partial E}{\partial s_i^a(n)}\equiv \delta_i^a(n)\ ,
\end{equation}
given $\delta_i^a(n+1)$. Computing the derivative 
we get
\begin{equation}
\frac{\partial E}{\partial s_i^a(n)} = 
\sum_j \frac{\partial E}{\partial s_j^b(n+1)}
\frac{\partial s_j^b(n+1)}{\partial s_i^a(n)} = 
\sum_j \delta_j^b(n+1)
\frac{\partial s_j^b(n+1)}{\partial s_i^a(n)}\ .
\label{eq:dEds}
\end{equation}
Next we compute the Jacobian $\frac{\partial s_j^b(n+1)}{\partial s_i^a(n)}$. For $i=j$ we have 
\begin{equation}
\frac{\partial s_i^b(n+1)}{\partial s_i^a(n)} = 
\frac{\partial s_i^b(n+1)}{\partial t_i^c(n)}
\frac{\partial t_i^c(n)}{\partial s_i^a(n)} = 
X_n^{bc}Z_{i,n}^{ca}\ ,
\end{equation}
and for $i\neq j$ we obtain
\begin{equation}
\frac{\partial s_j^b(n+1)}{\partial s_i^a(n)} = 
\frac{\partial s_j^b(n+1)}{\partial t_j^c(n)}
\frac{\partial t_j^c(n)}{\partial s_i^a(n)} = 
\frac{\partial s_j^b(n+1)}{\partial t_j^c(n)}
\frac{\partial t_j^c(n)}{\partial \theta_{n,j}}
\frac{\partial \theta_{n,j}}{\partial s_i^a(n)} =
X_n^{bc}\frac{\partial t_j^c(n)}{\partial \theta_{n,j}}
\gamma_n J_{ij}\delta^{az}\ .
\end{equation}
Thus Eq.~\eqref{eq:dEds} becomes 
\begin{equation}
\frac{\partial E}{\partial s_i^a(n)} = 
\delta_i^b(n+1)(X_nZ_{i,n})^{ba} + 
\sum_{j\neq i} \delta_j^b(n+1)X_n^{bc}
\frac{\partial t_j^c(n)}{\partial \theta_{n,j}}
\gamma_n J_{ij}\delta^{az}\ ,
\label{eq:dEds2}
\end{equation}
or, in vectorial form, as 
\begin{equation}
\boldsymbol{\delta}_i(n) = (X_nZ_{i,n})^T\boldsymbol{\delta}_i(n+1) + 
\Bigg[\gamma_n\sum_{j\neq i}J_{ij}\, 
\boldsymbol{\delta}_j(n+1) X_n 
\frac{\partial\boldsymbol{t}_j(n)}{\partial\theta_{n,j}}\Bigg]
\begin{pmatrix}
0\\0\\1
\end{pmatrix}\ ,
\label{eq:deltabackprop}
\end{equation}
which represents the backpropagation of the 
gradients from layer $n+1$ to layer $n$.

Next we move to the calculation of 
the gradients with respect to the 
variational parameters. The derivative 
w.r.t. $\beta_n$ yields
\begin{equation}
\frac{\partial E}{\partial \beta_n} = 
\sum_i \frac{\partial E}{\partial s_i^a(n+1)}
\frac{\partial s_i^a(n+1)}{\partial \beta_n} = 
\sum_i \delta_i^a(n+1)\frac{\partial s_i^a(n+1)}{\partial \beta_n}\ .
\end{equation}
Writing $\mathbf{s}_i(n+1)$ as 
\begin{equation}
\mathbf{s}_i(n+1) = 
\begin{pmatrix}
t_i^x(n) \\
t_i^y(n)\cos\beta_n - t_i^z(n)\sin\beta_n\\
t_i^y(n)\sin\beta_n + t_i^z(n)\cos\beta_n
\end{pmatrix}\ ,
\end{equation}
the derivative becomes 
\begin{equation}
\frac{\partial\mathbf{s}_i(n+1)}{\partial \beta_n} = 
\begin{pmatrix}
0 \\
-t_i^y(n)\sin\beta_n - t_i^z(n)\cos\beta_n\\
t_i^y(n)\cos\beta_n - t_i^z(n)\sin\beta_n
\end{pmatrix} =
\begin{pmatrix}
0 \\
-s_i^z(n+1)\\
s_i^y(n+1)
\end{pmatrix}\ ,
\end{equation}
and the gradient thus reads
\begin{equation}
\frac{\partial E}{\partial \beta_n} = 
\sum_i \big[-\delta_i^y(n+1)s_i^z(n+1) + 
\delta_i^z(n+1)s_i^y(n+1)\big]\ .
\end{equation}

\medskip

Next we compute the gradient w.r.t. $\gamma_n$
\begin{equation}
\frac{\partial E}{\partial \gamma_n} = 
\sum_i \frac{\partial E}{\partial t_i^a(n)}
\frac{\partial t_i^a(n)}{\partial \theta_{n,i}}
\frac{\partial \theta_{n,i}}{\partial\gamma_n}\ ,
\end{equation}
where 
\begin{equation}
\frac{\partial E}{\partial t_i^a(n)} = 
\frac{\partial E}{\partial s_i^b(n+1)}
\frac{\partial s_i^b(n+1)}{\partial t_i^a(n)} = 
\delta_i^b(n+1)X_n^{ba} \equiv 
w_i^a(n)
\ .
\end{equation}
Then writing $\mathbf{t}_i(n)$ as 
\begin{equation}
\mathbf{t}_i(n) = 
\begin{pmatrix}
s_i^x(n)\cos\theta_{n,i} - s_i^y(n)\sin\theta_{n,i}\\
s_i^x(n)\sin\theta_{n,i} + s_i^y(n)\cos\theta_{n,i}\\
s_i^z(n)
\end{pmatrix}\ ,
\end{equation}
the derivative becomes 
\begin{equation}
\frac{\partial \mathbf{t}_i(n)}{\partial \theta_{n,i}}= 
\begin{pmatrix}
-s_i^x(n)\sin\theta_{n,i} - s_i^y(n)\cos\theta_{n,i}\\
s_i^x(n)\cos\theta_{n,i} - s_i^y(n)\sin\theta_{n,i}\\
0
\end{pmatrix} = 
\begin{pmatrix}
-t_i^y(n)\\
t_i^x(n)\\
0
\end{pmatrix}\ .
\end{equation}
Noticing also that $\frac{\partial \theta_{n,i}}{\partial\gamma_n}=h_i^{(n)}$, the gradient 
becomes
\begin{equation}
\frac{\partial E}{\partial \gamma_n} = 
\sum_i \big[-w_i^x(n)t_i^y(n) + w_i^y(n)t_i^x(n)\big]h_i^{(n)}\ .
\end{equation}

Finally, the gradient w.r.t. $\alpha$ is 
\begin{equation}
\frac{\partial E}{\partial \alpha} = \sum_i
\frac{\partial E}{\partial s_i^a(0)}
\frac{\partial s_i^a(0)}{\partial \alpha} = 
\sum_i\big[-\delta_i^x(0)\sin\alpha + \delta_i^z(0)\cos\alpha\big]\ .
\end{equation}

For the Max-Cut problem, where the spins are initialized 
with random signs, $\mathbf{s}_i(0)=(\cos\alpha,0,\epsilon_i\sin\alpha)$, 
the only change is in the gradient with respect to $\alpha$, 
which becomes 
\begin{equation}
\frac{\partial E}{\partial\alpha} = \sum_i\big[-\delta_i^x(0)\sin\alpha + \epsilon_i\delta_i^z(0)\cos\alpha\big]\ ,
\end{equation}
while the backward recursion of Eq.~\eqref{eq:deltabackprop} 
and the gradients with respect to $\beta_n$ and $\gamma_n$ 
are unchanged, since the initialization enters only through 
$\mathbf{s}_i(0)$. 

\begin{figure}[h]
	\centering
	\includegraphics[width=0.95\textwidth]{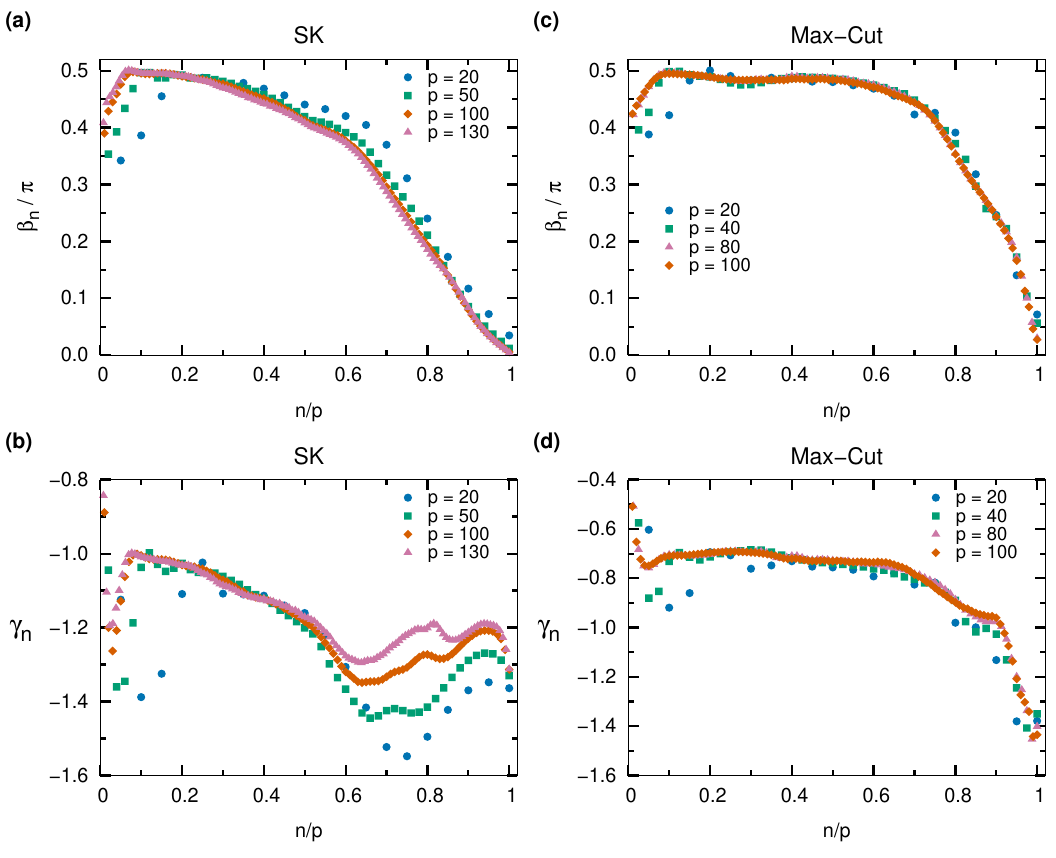}
    \caption{
    \justifying
\textbf{Optimized VIRAL schedules.}
{\bf(a)}, {\bf(b)} Optimized transverse angles 
$\beta_n$ (in units of $\pi$) and longitudinal 
angles $\gamma_n$ as a function of the rescaled 
layer index $n/p$ for the SK model, at depths 
$p=20,50,100,130$. 
{\bf(c)}, {\bf(d)} The same for Max-Cut on 
random 3-regular graphs, at depths $p=20,40,80,100$. 
With increasing depth the schedules collapse onto a 
limiting curve of $n/p$ in which, apart from the first 
few layers, $\beta_n$ decreases from $\approx\pi/2$ to 
$0$ while $\gamma_n$ remains negative and of order one, 
consistent with the optimal trajectories shown in 
Fig.~\ref{fig:figure3} of the main text.}
\label{fig:figureS1}
\end{figure}

\subsection{Training protocol}
\begin{figure}[h]
	\centering
	\includegraphics[width=0.6\textwidth]{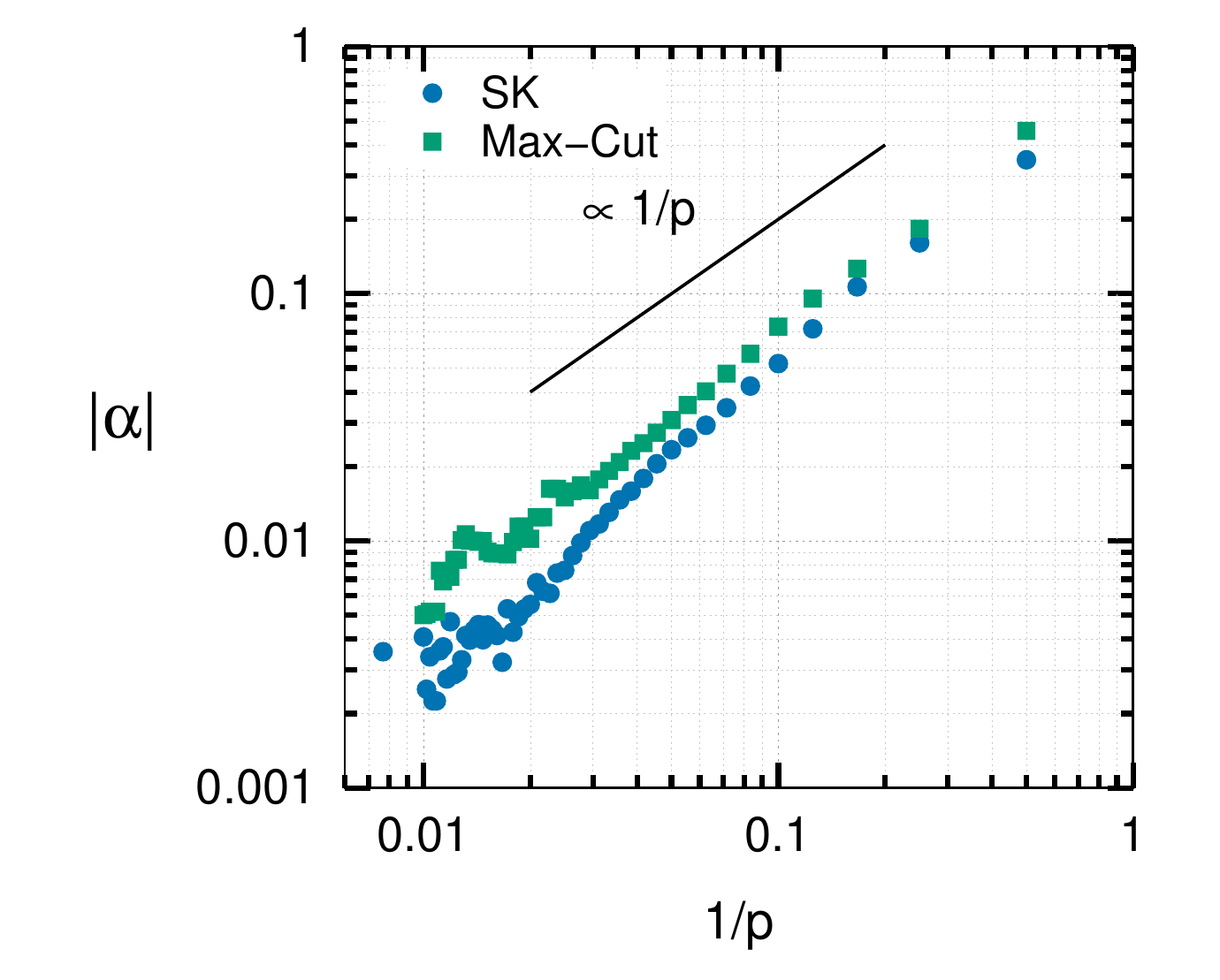}
    \caption{
    \justifying
\textbf{Optimized initial tilt.} 
Magnitude of the optimized initial angle $\alpha$ 
as a function of $1/p$, for the SK model (blue 
circles) and for Max-Cut on random 3-regular 
graphs (green squares). The optimal tilt vanishes 
roughly linearly with the depth, $|\alpha|\propto 1/p$ 
(black line), so that the initial state approaches 
the equatorial plane in the limit $p\to\infty$.}
\label{fig:figureS2}
\end{figure}

The variational parameters are trained with a curriculum 
over the circuit depth. The optimization starts at $p=2$, 
initialized with $\alpha=\pi/4$, $\beta_1=\pi/2$, $\beta_2=0$, 
and $\gamma_1=\gamma_2=-1.2$, and the depth is then 
increased in steps of two up to $p=130$, with the optimized 
schedule of each stage linearly interpolated to the next 
depth. In this way the optimization always starts near an 
optimum of the previous stage, which keeps the backpropagated 
gradients from exploding. At each stage, the exact 
gradients of the energy, averaged over the full batch of 
disorder instances, are computed by the backward recursion 
of Eq.~\eqref{eq:deltabackprop} and used to update the 
parameters with the Adam optimizer. The learning rate 
follows a cosine schedule, decreasing from $3\times10^{-3}$ 
to $10^{-5}$ and then periodically restarted to its 
maximal value with a doubling period. Each stage is stopped 
 when the best energy improves by less than $10^{-5}$ 
 over $50$ consecutive iterations, and the best parameters 
are carried over to the next depth. The same protocol is 
used for Max-Cut, where each instance consists of a random 
3-regular graph together with a quenched realization 
of the random signs $\{\epsilon_i\}$. 
The optimized schedules for both problems are shown in 
Fig.~\ref{fig:figureS1}, and the optimized initial tilt 
$\alpha$ in Fig.~\ref{fig:figureS2}. 

\section{Fixed points of the discrete map}
\label{sec:manybodykickedtopSI}

In this section we explore in detail the 
dynamics of the kicked tops evolving over 
time under the Hamiltonian in Eq.~\eqref{eq:manybodyH}, 
which we rewrite for clarity 
\begin{equation}
H(t) = \gamma(t) \sum_{i<j} J_{ij}\, s_i^z\, s_j^z 
    + \beta(t) \sum_{i=1}^{N} s_i^x\ ,
\label{eq:manybodyH_SI}
\end{equation}
There are two competing terms in this Hamiltonian: 
the precession non-linear term and the linear 
kick. The first one rotates each spin around the 
$z$ axis by an angle 
\begin{equation}
\theta_i = \gamma h_{i}^z=\gamma\sum_{j\neq i}J_{ij}\ s_j^z\ ,
\end{equation}
and this angle depends on the $z$ position of 
all spins $j\neq i$. The second linear term 
rotates all spins around the $x$ axis by a global 
angle $\beta$. 
The dynamics evolves according to the following 
discrete map 
\begin{equation}
\mathbf{s}_i(n+1) = R_x(\beta)R_z(\theta_i)\mathbf{s}_i(n)\ .
\label{eq:discreteMap_SI}
\end{equation}
The fixed points of the discrete map defined by 
Eq.~\eqref{eq:discreteMap_SI} are the solutions of 
\begin{equation}
\begin{aligned}
s_i^x &= s_i^x\cos(\theta_i) - s_i^y\sin(\theta_i)\ ,\\
s_i^y &= \cos(\beta)[s_i^x\sin(\theta_i) + s_i^y\cos(\theta_i)] - s_i^z\sin(\beta)\ ,\\
s_i^z &= \sin(\beta)[s_i^x\sin(\theta_i) + s_i^y\cos(\theta_i)] + s_i^z\cos(\beta)\ .
\end{aligned}
\label{eq:FixedPointMap_SI}
\end{equation}
The first equation can be rewritten as 
\begin{equation}
2\sin(\theta_i/2)[s_i^x\sin(\theta_i/2) + s_i^y\cos(\theta_i/2)] = 0\ .
\label{eq:twcasesSI}
\end{equation}
We have two cases depending on $\sin(\theta_i/2)=0$ 
or $\sin(\theta_i/2)\neq0$. 

\subsection{Case 1: $\sin(\theta_i/2)=0$. Equatorial fixed point}

In this case $s_i^z=0$, so the fixed point 
must be of the form $\mathbf{s}_i=(s_i^x,s_i^y,0)$. 
Since $R_z(\theta_i)=R(0)=I$, the map reduces to 
$\mathbf{s}_i(n+1) = R_x(\beta)\mathbf{s}_i(n)$, 
whose fixed points are $\mathbf{s}_i^*=(\pm1,0,0)$. 
These two equatorial fixed points exist for all 
values of $\gamma$ and $\beta$. 
Next we consider the fixed point $(1,0,0)$ (the analysis 
for $(-1,0,0)$ can be conducted in the same way). 
To study the stability of this fixed point we linearize 
around $\mathbf{s}_i^*$ by setting $\mathbf{s}_i=\mathbf{s}_i^* 
+\bm{\epsilon}_i$, 
where $\bm{\epsilon}_i=(0,\epsilon_{i,y}, \epsilon_{i,z})$. 
Applying $R_z$ and $R_x$ we obtain the evolution map 
for the perturbation 
\begin{equation}
\begin{aligned}
\epsilon^{(n+1)}_{i,y} &= \Big[\epsilon^{(n)}_{i,y} + 
\gamma\sum_{j\neq i}J_{ij}\epsilon^{(n)}_{j,z}\Big]\cos\beta - 
\epsilon^{(n)}_{i,z}\sin\beta\ ,\\
\epsilon^{(n+1)}_{i,z} &= \Big[\epsilon^{(n)}_{i,y} + 
\gamma\sum_{j\neq i}J_{ij}\epsilon^{(n)}_{j,z}\Big]\sin\beta + 
\epsilon^{(n)}_{i,z}\cos\beta\ .
\end{aligned}
\label{eq:pertMap_SI}
\end{equation}
We write the evolution equations~\eqref{eq:pertMap_SI} in 
matrix form by stacking all perturbations into a vector of 
size $2N$, denoted $(\bm{\epsilon}_y, \bm{\epsilon}_z)$, 
\begin{equation}
(\bm{\epsilon}_y, \bm{\epsilon}_z)^T \equiv 
(\epsilon_1^y,...,\epsilon_N^y, \epsilon_1^z,...,\epsilon_N^z)^T
\end{equation}
as 
\begin{equation}
\begin{pmatrix}
\bm{\epsilon}_y\\
\bm{\epsilon}_z
\end{pmatrix}^{(n+1)} = 
\begin{pmatrix}
\bm{A} & \bm{B}\\
\bm{C} & \bm{D}
\end{pmatrix}
\begin{pmatrix}
\bm{\epsilon}_y\\
\bm{\epsilon}_z
\end{pmatrix}^{(n)}\ ,
\end{equation}
where the $N\times N$ matrices $A,B,C$ and $D$ are 
defined as 
\begin{equation}
\begin{aligned}
\bm{A} &= \cos\beta\ \bm{I}\ ,\\
\bm{B} &= \gamma\cos\beta\ \bm{J} - \sin\beta\ \bm{I}\ ,\\
\bm{C} &= \sin\beta\ \bm{I}\ ,\\
\bm{D} &= \gamma\sin\beta\ \bm{J} + \cos\beta\ \bm{I}\ .
\end{aligned}
\end{equation}
A convenient feature of the matrix form is that all blocks 
are diagonalized by the same orthogonal transformation 
$O$ that diagonalizes the coupling matrix 
$\bm{J}=O\bm{\Lambda}_J O^T$. Thus, rotating the 
perturbation in the eigenbasis of $\bm{J}$  
$(\bm{\delta}_y,\bm{\delta}_z) = (O^T\bm{\epsilon}_y,O^T\bm{\epsilon}_z)$
we find 
\begin{equation}
\begin{pmatrix}
\bm{\delta}_y\\
\bm{\delta}_z
\end{pmatrix}^{(n+1)} = 
\begin{pmatrix}
\bm{A} & \bm{\Lambda}_B\\
\bm{C} & \bm{\Lambda}_D
\end{pmatrix}
\begin{pmatrix}
\bm{\delta}_y\\
\bm{\delta}_z
\end{pmatrix}^{(n)}\ ,
\end{equation}
where each block is now diagonal with 
\begin{equation}
\begin{aligned}
\bm{\Lambda}_B &= \gamma\cos\beta\ \bm{\Lambda}_J - 
\sin\beta\ \bm{I}\ ,\\
\bm{\Lambda}_D &= \gamma\sin\beta\ \bm{\Lambda}_J + \cos\beta\ \bm{I}\ .
\end{aligned}
\end{equation}
Thus, the problem factorizes into $N$ separate 
$2\times2$ subproblems of the form 
\begin{equation}
\begin{pmatrix}
\delta_k^y\\
\delta_k^z
\end{pmatrix}^{(n+1)} = 
\begin{pmatrix}
\cos\beta & \gamma\lambda_k\cos\beta - \sin\beta\\
\sin\beta & \gamma\lambda_k\sin\beta + \cos\beta
\end{pmatrix}
\begin{pmatrix}
\delta_k^y\\
\delta_k^z
\end{pmatrix}^{(n)}\ .
\label{eq:eigenpert_SI}
\end{equation}
Since the eigenvalues are complex conjugates 
with modulus 1, they can be written as 
$\mu_k^{\pm}=e^{\pm i\omega_k}$ where 
$\omega_k$ can be easily calculated from the 
trace of the matrix in Eq.~\eqref{eq:eigenpert_SI} 
and we obtain  
\begin{equation}
\omega_k = \cos^{-1}\Big(\cos\beta + \frac{\gamma\lambda_k}{2}\sin\beta\Big)\ .
\end{equation}
The equatorial fixed point is stable as long as all 
frequencies remain real (i.e. when the argument of 
the $\cos^{-1}$ is less than 1 in absolute value) 
yielding the stability condition
\begin{equation}
|2\cos\beta + \gamma\lambda_k\sin\beta|<2\ ,\ \ \forall\  k\ .
\end{equation}
Since the eigenvalues of the coupling matrix 
follow the Wigner semicircle law with support 
$\lambda\in[-2,2]$, the two edges of the spectrum 
control the onset of the instability. 

\medskip

For $\lambda=+2$, the stability condition gives 
\begin{equation}
-\cot(\beta/2)<\gamma<\tan(\beta/2)\ \ \ ,\ {\rm for}\ \ \lambda=+2\ .
\end{equation}
The lower and upper boundaries of this inequality 
give different types of instabilities. More precisely, 
at the upper boundary, when $\gamma\to \tan(\beta/2)$, 
the frequency $\omega\to 0$, while at the lower 
boundary, when $\gamma\to -\cot(\beta/2)$, the frequency 
$\omega\to \pi$. In the first case the instability is 
driven by the vanishing of a soft mode and the equatorial 
fixed point undergoes a pitchfork bifurcation into two 
polar fixed points with $s_i^z\neq0$; in the second case 
the instability is driven by a normal mode reaching the 
frequency $\omega=\pi$, signaling the emergence of a 
period-2 orbit in which the perturbation flips sign 
at every kick, $\boldsymbol{\delta} \to -\boldsymbol{\delta} \to \boldsymbol{\delta}$. 

\medskip

For $\lambda=-2$, the stability condition gives the 
reversed condition 
\begin{equation}
-\tan(\beta/2)<\gamma<\cot(\beta/2)\ \ \ ,\ {\rm for}\ \ \lambda=-2\ ,
\end{equation}
where the role of the boundaries is inverted: 
the upper boundary ($\gamma \to \cot(\beta/2$) 
corresponds to $\omega \to \pi$ and drives a 
period-2 instability, while the lower boundary 
($\gamma \to -\tan(\beta/2$) corresponds to $\omega \to 0$ 
and drives a pitchfork bifurcation. 
The full stable region is the intersection of these 
two conditions, bounded by all four curves 
$\gamma = \pm\tan(\beta/2)$ and $\gamma = \pm\cot(\beta/2)$. 
These curves meet at two multicritical points 
$(\beta, \gamma) = (\pi/2, \pm 1)$ where both instabilities 
occur simultaneously, as shown in Fig.~\ref{fig:figure3}a 
in the main text (for Max-Cut the spectral support changes 
and the multicritical points move to $(\pi/2,\pm1/\sqrt2)$, 
as explained in Sec.~\ref{sec:kickedtops}).

\medskip

\begin{figure}[h]
	\centering
	\includegraphics[width=0.6\textwidth]{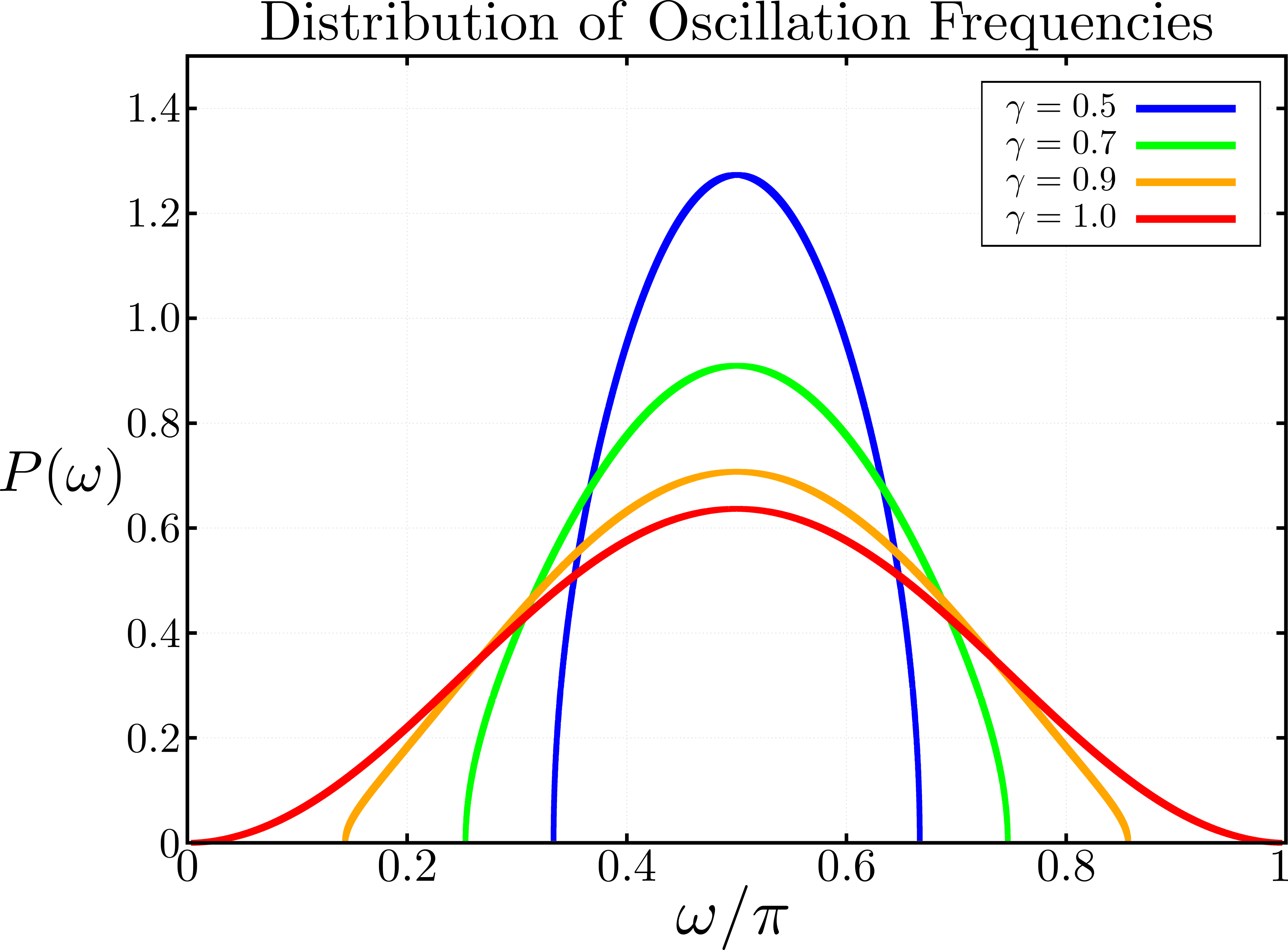}
    \caption{
    \justifying
   \textbf{Distribution of normal mode frequencies.} 
    Probability density $P(\omega)$ of the frequencies of 
    normal modes around the fixed point 
    $\mathbf{s}_i^*=(1,0,0)$    
    for different values of $\gamma$ (with $\beta=\pi/2$). 
    For small $\gamma$ (e.g., $\gamma=0.5$, blue curve), 
    the frequencies are concentrated in a narrow band around $\omega=\pi/2$. As $\gamma$ increases towards the critical 
    value $\gamma_c=1$, the width of the distribution expands 
    and, at the critical point $\gamma=1$ (red curve), the support 
    of $P(\omega)$ spans the entire range $\omega \in [0, \pi]$. 
    The non-zero density at $\omega \to 0$ indicates the emergence 
    of soft modes and at $\omega \to \pi$ the emergence of period-2 orbits. 
    }
\label{fig:figureS3}
\end{figure}

The distribution of the oscillation frequencies 
$P(\omega)$ (i.e. the density of states) can be obtained 
simply as $P(\omega) = \rho(\lambda)\Big|\frac{d\lambda}{d\omega}\Big|$, 
where $\rho(\lambda)=\frac{1}{2\pi}\sqrt{4-\lambda^2}$ is 
the Wigner semicircle function. For $\beta=\pi/2$ the density 
of states has the simple form shown in Fig.~\ref{fig:figureS3}. 
\begin{equation}
P(\omega) = \frac{2}{\pi\gamma^2}|\sin(\omega)|\sqrt{\gamma^2-\cos^2(\omega)}\ \ \  {\rm for}\ \ \omega\in[\omega_{\min},\omega_{\max}]\ ,
\label{eq:p_omegaSI}
\end{equation} 
with the lowest and highest frequencies given by 
\begin{equation}
\omega_{\min} = \cos^{-1}(\gamma)\ ,\ \ \ 
\omega_{\max} = \cos^{-1}(-\gamma)\ .
\end{equation}
In particular, for $\gamma\to\gamma_c$, we find 
$\omega_{\min}\to 0$ as 
\begin{equation}
\omega_{min}\sim \sqrt{\gamma_c-\gamma}\ \ \ {\rm for}\ \ \gamma\to\gamma_c^-\ .
\end{equation}

\subsection{Case 2: $\sin(\theta_i)\neq0$. Polar fixed points}
In this case, the fixed point equations are the 
solution of  
\begin{eqnarray}
s_i^x&=&\frac{\tan(\beta/2)}{\tan(\theta_i/2)}\ s_i^z, \nonumber \\
s_i^y&=&-\tan(\beta/2)\ s_i^z, \nonumber \\
s_i^z&=&\frac{\sin(\theta_i/2)}{\sqrt{\tan^2(\beta/2)+\sin^2(\theta_i/2)}}\ .
\label{eq:polarfixpointSI}
\end{eqnarray}

\medskip

{\bf $\beta=0$. Ising-like fixed points}

\medskip

It is easy to see that when $\beta=0$, the polar 
fixed points are located exactly at the poles, 
i.e. 
\begin{equation}
\mathbf{s}_i^*=(0,0,{\rm sign}(\theta_i^*))\ ,    
\end{equation}
where $\theta_i^*$ is given by 
\begin{equation}
\theta_i^* = \gamma\sum_{j\neq i}J_{ij}(s_j^z)^*\equiv 
\gamma h_i^*\ .
\end{equation}
There are exactly $2^N$ of such fixed points, 
corresponding to all possible Ising-like configurations. 
One of this fixed points is precisely the ground state 
of the SK Hamiltonian that the optimal protocol 
seeks to find. 

A perturbation around one of these fixed points 
is, at linear order, completely confined to the 
$xy$-plane, i.e. $\bm{\epsilon}_i=(\epsilon_{i,x}, \epsilon_{i,y},0)$. 
Substituting in Eq.~\eqref{eq:polarfixpointSI} 
and keeping only term up to first order, the 
problem decouples into $N$ independent $2\times2$ 
subproblems, one for each site $i$ as
\begin{equation}
\begin{pmatrix}
\epsilon_i^x\\
\epsilon_i^y
\end{pmatrix}^{(n+1)} = 
\begin{pmatrix}
\cos\theta_i^* & -\sin\theta_i^*\\
\sin\theta_i^* & \cos\theta_i^*
\end{pmatrix}
\begin{pmatrix}
\epsilon_i^x\\
\epsilon_i^y
\end{pmatrix}^{(n)}\ ,
\label{eq:polarpert_SI}
\end{equation}
and the eigenvalues of the Jacobian are simply 
$\mu_i^{\pm} = e^{\pm i\theta_i^*}$. Therefore the frequencies 
of the normal modes are simply 
\begin{equation}
\omega_i =  \theta_i^* = \gamma h_i^*\ .
\end{equation}
Since the eigenvalues lie on the unit circle 
for any value of $\gamma h_i^*$, the polar 
fixed points, at $\beta=0$, are always stable. 

\medskip

{\bf $\beta\ll1$. Stability of the Ising-like fixed points}

\medskip

When $\beta\neq 0$ but small, the kick on the x-axis can 
be treated as a perturbation. This perturbation will 
change the eigenvalues $\mu_i$, but since the dynamics 
is Hamiltonian, the determinant of the Jacobian must 
be $1$, and thus the perturbed eigenvalues must still 
satisfy $\mu_i^+\mu_i^-=1$. If the unperturbed eigenvalues 
are not on the real axis, the perturbation will 
cause just a frequency shift $\theta_i^*\to \theta_i^*+\beta$. 
However if the unperturbed eigenvalues are degenerate 
so that $\mu_i^+=\mu_i^-=1$ (i.e. $\omega_i=0$) 
or $\mu_i^+=\mu_i^-=-1$ (i.e. $\omega_i=\pi$), then 
the perturbation will push one eigenvalue outside 
the unit circle causing the spin to leave the z-axis. 
Thus, for the Ising fixed point to be stable against 
small $\beta$ kicks, the following condition must hold 
\begin{equation}
0<|\gamma h_i^*|<\pi, \ \ \ \forall i\ \ \ {\rm Stability\ of\ the\ Ising\  FP}\ .
\end{equation}
The right side of this condition sets the maximum 
value of the local field to 
\begin{equation}
|h^*_{max}|<\frac{\pi}{|\gamma|}\ ,
\end{equation}
which, using the last layer value $|\gamma_p|\sim~1.3$ 
of the optimized protocol, gives $|h^*_{max}|\lesssim 2.4$. 
A spin with a local field near $|h^*_{max}|$ 
will be pushed away from the z-axis by a small $\beta$ 
kick, causing an energy increase of order $|h^*_{max}|$ 
above the Parisi value. 
Similarly, a spin with a local field very close 
to zero will also drift away from the z-axis in 
presence of a small kick. Indeed, if $h_i^*\sim0$ 
and $\beta$ is small, we can write the z-component 
of the spin as 
\begin{equation}
s_i^z\sim\frac{\gamma h_i^*}{\sqrt{\beta^2+(\gamma h_i^*)^2}}\sim0\ \ \ 
{\rm for}\ \ \ \ |\gamma h_i^*|\ll\beta,
\end{equation}
i.e. all spins with a local field $|h_i^*|\lesssim \beta/\gamma $ 
will be dragged to the equator by the transverse kick. 
Although the direct contribution to the energy is 
negligible (since its $h_i^*\sim0$), once the spin 
leaves the z-axis it lowers the local fields 
on its neighbors. This shift might push one of 
those local fields closer to 0, causing the 
neighboring spin to leave the z-axis, which 
in turn lowers the local fields on the neighbors 
even more, potentially triggering an avalanche 
that destabilizes the entire Ising fixed point. 
The number of spins that collapse to the equator 
when $\beta\neq0$ depends on the behavior of the 
distribution $P(h)$ near $h\sim0$. 
In the exact Parisi solution of the SK model at $T=0$, 
the probability distribution $P(h)$ vanishes linearly 
near $h=0$ as $P(h)\sim c|h|$. Therefore the fraction 
of spins dragged to the equator by a transverse 
kick of $O(\beta)$ is 
\begin{equation}
f_{z\to xy} = \int_{-\beta/\gamma}^{\beta/\gamma}P(h)dh \sim 
c\left(\frac{\beta}{\gamma}\right)^2\ ,
\label{eq:fraction_miss}
\end{equation}
so it's order $O(\beta^2)$. From the point of view of 
the algorithm, this means that as $\beta$ approaches zero 
in the final layers, the fraction of spins still near 
the equator is only $f\sim O(\beta^2)$. 
Since a given equatorial spin has a local field $h_i\sim O(\beta)$, 
the maximum energy contribution this spin could provide 
if it fully polarizes on the z-axis is $e_{max}=O(\beta)$. 
If it fails to polarize, instead, it would provide 
an energy contribution of only $e_{min}=O(\beta^2)$, since its 
local field is $O(\beta)$ and its z-components is also $O(\beta)$. 
Thus, the energy lost if it doesn't polarize is 
at most $O(\beta)$. So, by terminating the algorithm 
at a small nonzero $\beta$, the total energy lost is only 
$O(\beta^3)$, i.e. 
\begin{equation}
E(\beta)=E(0)+O(\beta^3)\ \ \ {\rm for}\ \ \ \beta\to 0\ .
\label{eq:energy_miss}
\end{equation}
The results in Eqs.~\eqref{eq:fraction_miss} 
and~\eqref{eq:energy_miss} also guarantee that 
the final thresholding step, $\sigma_i=\mathrm{sign}(s_i^z)$, 
does not compromise the ground state energy produced 
by the backbone of fully polarized spins. 
In the practical case, however, the $P(h)$ is never truly 
zero at $h=0$. In this case the fraction of spins near 
the equator is $f\sim O(\beta)$ and the energy error 
is $E(\beta)-E(0)=O(\beta^2)$. This quadratic scaling 
implies that to reach a fixed accuracy more layers 
are needed than in the ideal cubic scaling case.

\section{Reduction to a single-spin effective problem}

\subsection{Effective single-spin dynamics}
\label{sec:viralDMFT}

Consider a set of spins $s^{\mu}_i$ where 
$\mu$ denotes the component of the spin, 
$\mu=x,y,z$, and $i$ labels the site. 
They all evolve under the Landau-Lifshitz 
dynamics 
\begin{equation}
\frac{d \mathbf{s}_i}{dt}= \mathbf{h}_i(t) \times \mathbf{s}_i\ ,
\end{equation}
or by components 
\begin{equation}
\frac{d s_i^{\alpha}}{dt}= \epsilon^{\alpha \mu \nu} 
h_i^\mu(t) s_i^\nu\ ,
\end{equation}
where $\epsilon^{\alpha \mu \nu}$ is 
the Levi-Civita symbol and $h_i^\mu(t)$ is 
the local effective field expressed as
\begin{equation}
h_i^\mu(t) = 
\begin{cases} 
\sum_{j\neq i} \gamma(t) J_{ij}s^z_j(t) & \text{if } \mu = z, \\ 
\sum_{n=1}^p \beta_n\delta(t-n) & \text{if } \mu = x, \\
0 & \text{if } \mu = y\ ,
\end{cases}
\end{equation}
where $\gamma(t)$ is a piecewise-constant 
function of time, $\gamma(t) = \sum_{n=1}^{p} \gamma_n\,{\bf 1}{[n-1 \leq t < n]}$. 
Next consider the density function $\rho(\mathbf{s},t)$ 
representing the probability of finding the spin 
configuration $\mathbf{s}$ at time $t$. The time 
evolution of $\rho(\mathbf{s},t)$ is described by 
the Liouville equation
\begin{eqnarray}
\partial_t \rho = \sum_i (h_i\times s_i)\cdot \nabla_{s_i}\rho = 
-i\mathcal{L}\rho\ ,
\end{eqnarray}
where $\mathcal{L}$ is the Liouvillian operator. 
The path integral of the propagator $P(s_f,t_f|s_0,0)$ 
can be derived straightforwardly by: {\it (i)} breaking 
down the total time interval into $M$ small 
$\delta t = t_f/M$ , {\it (ii)} expressing the evolution 
operator as 
\begin{eqnarray}
e^{-i\mathcal{L}t_f}= e^{-i\mathcal{L}\delta t}\cdots e^{-i\mathcal{L}\delta t}\ ,
\end{eqnarray}
and {\it (iii)} inserting the identity at each time slice, 
thus obtaining 
\begin{equation}
P_J(\mathbf{s}_f,t_f|\mathbf{s}_0,0)=\int \mathcal{D}r(t) \int_{s(0)=s_0}^{s(t_f)=s_f} \mathcal{D}s(t) \exp{ \left[ i\sum_i \int_0^{t_f} {\rm d}t\ r_i^\alpha \Big(\dot{s}_i^{\alpha}- \epsilon^{\alpha \beta \gamma} h_i^\beta(t) s_i^\gamma\Big)\right] }.
\end{equation}
The part of the exponent which depends on $J_{ij}$ is 
\begin{equation}
-i\sum_{i\neq j}J_{ij}\int_0^{t_f} 
dt\ \gamma(t)\ r_i^\alpha(t) \epsilon^{\alpha z \gamma} 
s_i^\gamma(t) s^z_j(t)\equiv 
-i\sum_{i\neq j}J_{ij}\int_0^{t_f}dt\ \gamma(t)\ v_i(t)s^z_j(t) \equiv 
-i\sum_{i\neq j}J_{ij}A_{ij}
\end{equation}
where we defined 
\begin{equation}
\begin{aligned}
v_i(t) &\equiv r_i^\alpha(t) \epsilon^{\alpha z \gamma}s_i^\gamma(t)\ ,\\
A_{ij} &\equiv \int_0^{t_f}dt\ \gamma(t)\ v_i(t)s^z_j(t)\ .
\end{aligned}    
\end{equation}
Averaging $e^{-i\sum_{i\neq j}J_{ij}A_{ij}}$ over $J_{ij}$ 
we obtain 
\begin{equation}
\left\langle e^{-i\sum_{i\neq j}J_{ij}A_{ij}}\right\rangle_J = 
\int \Bigg(\prod_{i<j}P(J_{ij})dJ_{ij}\Bigg)e^{-i\sum_{i\neq j}J_{ij}A_{ij}}=
\exp\left[-\frac{1}{2N}\sum_{i< j}(A_{ij}+A_{ji})^2\right]\ .
\end{equation}
Then expanding the square in the sum we get 
\begin{equation}
\begin{aligned}
-\frac{1}{2N}\sum_{i< j}(A_{ij}+A_{ji})^2 = 
-\frac{1}{2N}\sum_{i\neq j}(A_{ij}^2 + A_{ji}A_{ij}) &=  
-\frac{1}{2}\sum_i\int dt dt'\ \gamma(t)\gamma(t')\ v_i(t)\Sigma(t,t')v_i(t')\ -\\
&=-\frac{N}{2}\int dt dt'\ \gamma(t)\gamma(t')\ R(t,t')R(t',t)
\end{aligned}
\label{eq:effAction}
\end{equation}
where 
\begin{equation}
\begin{aligned}
\Sigma(t,t')&=\frac{1}{N} \sum_j s^z_j(t)s^z_j(t')\ ,\\
R(t,t')&=\frac{1}{N} \sum_j v_j(t)s^z_j(t')\ ,   
\end{aligned}
\label{eq:Sigma_definition}
\end{equation}
are the spin-spin correlation function and 
the response function, respectively. 
Physically, $R(t,t')$ measures how much the 
$z$-component of the spin at time $t'$ changes 
if we apply a perturbation at time $t$. Therefore, 
the response function $R(t,t')$ must obey causality, 
meaning that $R(t,t')\propto\Theta(t'-t)$, so it must 
be identically zero for $t>t'$. As a consequence, the 
product $R(t,t')R(t',t)\propto\Theta(t'-t)\Theta(t-t')=0$, 
and thus we can drop the last integral in Eq.~\eqref{eq:effAction}. 

Next we perform a Hubbard-Stratonovich transformation. 
Setting $K(t,t') = \gamma(t)\gamma(t')\Sigma(t,t')$ 
we write 
\begin{equation}
\exp\left[-\frac{1}{2}\sum_i\int dtdt'\ v_i(t)K(t,t')v_i(t')\right] = 
\int \mathcal{D}g^z\ e^{-S[g^z]}\ e^{i\sum_i\int dt\ g^z(t)v_i(t)}\ ,
\end{equation}
where $S[g^z]$ is a Gaussian action defined as 
\begin{equation}
\begin{aligned}
S[g^z] = \frac{1}{2}\int dtdt'\ g^z(t)K^{-1}(t,t')g^z(t')\ ,\ \ \ \ 
K^{-1}(t,t') &= \frac{\Sigma^{-1}(t,t')}{\gamma(t)\gamma(t')}
\end{aligned}
\end{equation}

Defining the stochastic magnetic field 
$g^\mu(t)$ as
\begin{equation}
g^\mu(t) = 
\begin{cases}
g^z(t) & \text{if } \mu = z, \\ 
\sum_{n=1}^p \beta_n\delta(t-n) & \text{if } \mu = x, \\
0 & \text{if } \mu = y\ ,
\end{cases}
\end{equation}
we can write the disorder averaged propagator as 
$\langle P_J(\mathbf{s}_f,t_f|\mathbf{s}_0,0)\rangle_J\equiv 
 P(\mathbf{s}_f,t_f|\mathbf{s}_0,0)$ as 
\begin{equation}
P(\mathbf{s}_f,t_f|\mathbf{s}_0,0) =
\int \mathcal{D}g^z \mathcal{D}r \mathcal{D}s\ 
\exp\left[-\frac{1}{2}\int dt dt'\  
\frac{g^z(t) \Sigma^{-1}(t,t') g^z(t')}{\gamma(t)\gamma(t')} + i \sum_i \int_0^{t_f} {\rm d}t\ r_i^\alpha \big(\dot{s}_i^{\alpha}- \epsilon^{\alpha \beta \gamma} g^\beta(t) s_i^\gamma\big)  \right]\ .
\label{eq:averagepropagator1}
\end{equation}
Integrating out the response fields $r^\alpha_i$ 
generates delta functions that constrain each spin 
to obey an effective LLG equation and we get 
\begin{equation}
P(\mathbf{s}_f,t_f|\mathbf{s}_0,0) =
\int \mathcal{D}g^z\mathcal{D}s\ e^{-S[g^z]}\ 
\prod_{i,t}\delta\big[\dot{s}_i^{\alpha}- \epsilon^{\alpha \beta \gamma} g^\beta(t) s_i^\gamma\big]\ .
\label{eq:averagepropagator2}
\end{equation}
Now, all spins have the same initial condition 
and obey the same equation of motion for each 
realization of $g^z$, so the problem reduces to 
a single-spin problem with propagator 
\begin{equation}
P(s_f,t_f|s_0,0) =
\int \mathcal{D}g^z\mathcal{D}s\ e^{-S[g^z]}\ \prod_t
\delta\big[\dot{s}^{\alpha}- \epsilon^{\alpha \beta \gamma} g^\beta(t) s^\gamma\big]\ .
\label{eq:averagepropagator3}
\end{equation}
The effective single-spin LLG dynamics is given by 
\begin{equation}
\partial_t s^{\alpha}= \epsilon^{\alpha \beta \gamma} g^\beta(t) s^\gamma\ ,
\label{eq:singlespineff}
\end{equation}
 with 
\begin{equation}
\langle\langle g^z(t)g^z(t')\rangle\rangle = 
\gamma(t)\gamma(t')\Sigma(t,t') =
\gamma(t)\gamma(t')\langle\langle s^z(t)s^z(t')\rangle\rangle\ , 
\end{equation}
where the average $\langle\langle \cdot \rangle\rangle$ 
is taken using the distribution~\eqref{eq:averagepropagator3}. 
Since $\Sigma$ depends on the trajectories, and the 
trajectories depend on $g^z$, and $g^z$ is drawn from 
a Gaussian with covariance $\gamma(t)\gamma(t')\Sigma(t,t')$, 
the whole thing must be solved self-consistently. 
This is a subtle point that we would like to clarify. 
Looking back at the definition of $\Sigma$ in Eq.~\eqref{eq:Sigma_definition}, we can see that 
$\Sigma$ is not a fixed external object: it is 
a functional of all $N$ spin trajectories. 
The disorder average converted the quenched 
random interactions $J_{ij}$ into a self-interaction 
non-local in time mediated by $\Sigma(t,t')$, 
and the Hubbard-Stratonovich transformation 
introduced the auxiliary field $g^z(t)$. 
The point is that the Gaussian weight of $g^z$ 
in Eq.~\eqref{eq:averagepropagator3} involves 
$\Sigma^{-1}$, which depends itself on the same 
trajectories that $g^z$ drives. The measure $P(s,g)$ 
\begin{equation}
P(s,g^z)\propto
\exp\left[-\frac{1}{2}\int dt dt'\  
\frac{g^z(t) \Sigma^{-1}(t,t') g^z(t')}{\gamma(t)\gamma(t')}
\right]
\prod_t\delta \left[ \dot{s}^{\alpha}- \epsilon^{\alpha \beta \gamma} g^\beta(t) s^\gamma \right] .
\label{eq:selfconsistentMeasure}
\end{equation}
is therefore defined only at a self-consistent 
fixed point solution $\Sigma^*$.

This self-consistency condition is obtained as 
follows. Since all spins share the same initial 
condition $s_i(0) = (\cos\alpha, 0, \sin\alpha)$ 
for all $i$, and since after the Hubbard-Stratonovich 
transformation every spin $i$ obeys the same equation 
of motion, $\dot{s}_i^\alpha = \epsilon^{\alpha\beta\gamma}\,g^\beta(t)\,s_i^\gamma$, 
all spins follow exactly the same trajectory for each 
realization of $g^z$, i.e. 
\begin{equation}
s_i(t) = s(t;\, g^z)\quad \forall\ i\ .
\label{eq:identicalLLG}
\end{equation}
Substituting into Eq.~\eqref{eq:Sigma_definition}, 
for each realization of $g^z$, the correlator $\Sigma$ 
is exactly rank-1:
\begin{equation}
\Sigma(t,t') = s^z(t; g^z) s^z(t'; g^z)\ .
\label{eq:Sigma_rank1_sample}
\end{equation}
Averaging over $g^z$, the self-consistency condition 
therefore reads
\begin{equation}
\Sigma^*(t,t') = \bigl\langle s^z(t;\,g^z)\,s^z(t';\,g^z)\bigr\rangle_{g^z},
\qquad {\rm with}\ \ \  g^z \sim \mathcal{N}\left(0,\,\gamma_t\gamma_{t'}\Sigma^*_{tt'}\right)\ ,
\label{eq:self_consistency}
\end{equation}
which is a fixed-point equation for the function 
$\Sigma^*(t,t')$. In words, the meaning of this 
self-consistent equation is: the stochastic magnetic 
field $g^z(t)$ that drives the spin is a Gaussian 
process with covariance $\gamma(t)\gamma(t')\Sigma^*(t,t')$, 
and $\Sigma^*$ must equal the two-time correlator
produced by the spin evolving under that same 
stochastic field. 

In principle, Eq.~\eqref{eq:self_consistency} can be 
solved iteratively from a starting point $\Sigma^{(0)}(t,t')$. 
The natural choice is to initialize the iteration scheme 
by setting $g^z(t)=0$ at the beginning. 
When $g^z = 0$, the effective LLG dynamics is fully 
deterministic and has a unique solution $s_z^{(0)}(t)$ 
given the initial condition $s_z^{(0)}(0)=\sin\alpha$. 
This gives the zero order correlator as
\begin{equation}
\Sigma^{(0)}(t,t') = s^{(0)}_z(t)s^{(0)}_z(t')\ ,
\label{eq:Sigma_init}
\end{equation}
which is exactly rank-1. 
Given $\Sigma^{(n)}$ at iteration step $n$, at the 
subsequent iteration $n+1$ we proceed as follows: 
\begin{itemize}
\item (i) draw $g^z \sim \mathcal{N}(0,\gamma_t\gamma_{t'}\Sigma^{(n)}_{tt'})$; 
\item (ii) integrate Eq.~\eqref{eq:singlespineff} to obtain $s^z(t;g^z)$; 
\item (iii) set $\Sigma^{(n+1)}(t,t') = \langle s^z(t)s^z(t')\rangle_{g^z}$, and repeat until convergence. 
The fixed point $\Sigma^*$ then defines the effective 
measure $P(s,g)$ in Eq.~\eqref{eq:selfconsistentMeasure}. 
\end{itemize}

Note that the rank-1 property holds exactly 
within each sample of $g^z$. After averaging, 
$\Sigma^*(t,t') = \langle\langle s^z(t)s^z(t')\rangle\rangle$ need not be rank-1 as a matrix in $(t,t')$. 
The claim that $\Sigma^*$ remains approximately 
rank-1 throughout the dynamics, as seen numerically 
in Fig.~\ref{fig:figure4}d, is the key difference 
with the quantum case. In QAOA, the qubits have $(\sigma^z)^2=1$, which implies the initial variance is 1.

\subsection{How to compute the average energy}
\label{sec:pathIntegralSI}

The quantity of interest is the disorder-averaged 
energy of the final state:
\begin{equation}
E=\Big\langle \sum_{i<j}J_{ij}s^z_i(t_f)s_j^z(t_f)\Big\rangle_{\rho_f}\ ,
\label{eq:energy_rhof}    
\end{equation}
where ${\rho_f}\equiv \rho(\mathbf{s}_f,t)$ is 
given by 
\begin{equation}
\rho(\mathbf{s}_f,t) = \int 
\langle P(\mathbf{s}_f,t_f|\mathbf{s}_0,0)\rangle_J\ 
\rho(\mathbf{s}_0,0)\ d\mathbf{s}_0\ .
\end{equation}
Here we assume that at time $t=0$ all spins 
lie in the $xz$-plane and point in the same 
direction specified by the angle $\alpha$, 
which is itself a variational parameter. 
In other words $\rho(\mathbf{s}_0,0)$ is 
a product of delta-functions 
\begin{equation}
\rho(\mathbf{s}_0,0) = \prod_{i=1}^N\delta(s_{0i}^x-\cos\alpha)\delta(s_{0i}^y)\delta(s_{0i}^z-\sin\alpha)\ .
\end{equation}
We extract the energy by differentiating 
the generating functional with respect to 
a source $\eta$ 
\begin{equation}
E= \frac{\partial}{\partial\eta}\Bigg|_{\eta=0}\mathbb{E}_J\left\{\int \rho(\mathbf{s}_0,0)\ d\mathbf{s}_0
\int \mathcal{D}r \int_{s_0}^{s_f} \mathcal{D}s \exp{ \left[ i \int_0^{t_f} {\rm d}t\ \sum_ir_i^\alpha \Big(\dot{s}_i^{\alpha}- \epsilon^{\alpha \beta \gamma} h_i^\beta(t) s_i^\gamma\Big) -i\frac{\eta}{2} \sum_{ij}J_{ij}s^z_is_j^z \delta(t-t_f) \right] }
\right\}\ .
\end{equation}
This description is convenient because it allows 
us to take easily the average over the quenched disorder.  
Note that this is exactly what is needed, because 
we want to come up with controls $\gamma(t)$ and 
$\beta(t)$ that do not depend on the exact realization 
of the $J_{ij}$'s and as such minimize the average energy 
of the ensemble. 
There are only two terms in the exponent that depend 
on $J_{ij}$, namely the term from the local field 
$h_i^{z}(t)$ and the boundary term which comes with a 
factor of $\eta$
\begin{equation}
-i\sum_{i\neq j} J_{ij}A_{ij} + 
\frac{\eta}{2}\sum_{i\neq j} J_{ij}\ s^z_i(t)\ s_j^z(t)\ ,
\end{equation}
where again we introduced the shorthand 
$A_{ij}\equiv \int_0^{t_f}dt\ \gamma(t)\ v_i(t)s^z_j(t)$ 
with $v_i(t)\equiv r_i^\alpha(t) \epsilon^{\alpha z \gamma} s_i^\gamma(t)$. 
Performing the average over $J_{ij}$ results in an 
exponent 
\begin{equation}
\frac{1}{2N}\sum_{i<j}\Big[-i(A_{ij} + A_{ji}) + 
\eta s^z_i(t)\ s_j^z(t)\Big]^2
\end{equation}
Expanding the square and summing over $j$ as 
in Eq.~\eqref{eq:effAction} generates 
$\Sigma(t,t')=\frac{1}{N}\sum_j s_j^z(t)s_j^z(t')$. 
We first analyze the $\eta^0$ term: this gives
the effective action already derived in Eq.~\eqref{eq:effAction}
\begin{equation} 
-\frac{1}{2}\sum_i\int dt dt'\ \gamma(t)\gamma(t')\ v_i(t)\ \Sigma(t,t')\ v_i(t')\ .
\label{eq:zeroordereta}
\end{equation}
The term linear in $\eta$ is the only one contributing 
to $E=\partial_\eta|_{\eta=0}$ and reads
\begin{equation}
-\frac{i\eta}{N}\sum_{i\neq j} A_{ij}\ s^z_i(t_f)\ s^z_j(t_f) =
-i\eta\sum_{i}\int_0^{t_f}dt\ \gamma(t)\ v_i(t)\ 
 \Sigma(t,t_f)\ s^z_i(t_f)\  .
\label{eq:firstordereta}
\end{equation}
Then, differentiating at $\eta = 0$ we obtain the following 
expression for the disorder-averaged energy 
\begin{equation}
E=-i\int \mathcal{D}r\mathcal{D}s\mathcal{D}g^z\ 
e^{-S[g^z]}\ \Bigg[\sum_{i}\int_0^{t_f}dt\ \gamma(t)\ v_i(t)\ \Sigma(t,t_f)\ s^z_i(t_f)\Bigg]\ 
\exp\Bigg[i\sum_i \int_0^{t_f} dt'\ r_i^\alpha \Big(\dot{s}_i^{\alpha}- \epsilon^{\alpha \beta \gamma} g^\beta(t') s_i^\gamma\Big) \Bigg]\ . 
\end{equation}
As shown before, after the Hubbard-Stratonovich 
transformation the path integral factorizes over 
sites so the $\sum_i$ inside the path integral 
yields a factor $N$ and we are left with 
\begin{equation}
\frac{E}{N}=-i\int \mathcal{D}r\mathcal{D}s\mathcal{D}g^z\ 
e^{-S[g^z]}\ \int_0^{t_f}dt\ \gamma(t)\ v(t)\ \Sigma(t,t_f)\ s^z(t_f)\ e^{i\Phi(s,r,g)}\ ,
\end{equation}
where 
\begin{equation}
\Phi(s,r,g) \equiv  \int_0^{t_f} dt'\ r^\alpha \Big(\dot{s}^{\alpha}- \epsilon^{\alpha \beta \gamma} g^\beta(t') s^\gamma\Big)\ .
\end{equation}
Now we observe that the factor $v(t)\equiv r^\alpha(t) \epsilon^{\alpha z \gamma} s^\gamma(t)$ inside the 
path integral can be written as the functional derivative 
of the exponent with respect to $g^z(t)$. Indeed since 
$g^z(t)$ appears in the $\Phi$ only through 
$-r^\alpha\epsilon^{\alpha z \gamma} g^z s^\gamma=-vg^z$ 
we can write 
\begin{equation}
v(t)\ e^{i\Phi(s,r,g)} = i\frac{\delta }{\delta g^z(t)}\ e^{i\Phi(s,r,g)}
\end{equation}
and we get
\begin{equation}
\frac{E}{N}=\int \mathcal{D}r\mathcal{D}s\mathcal{D}g^z\ 
e^{-S[g^z]}\ \int_0^{t_f}dt\ \gamma(t)\  \Sigma(t,t_f)\ s^z(t_f)\frac{\delta}{\delta g^z(t)}\  
e^{i\Phi(s,r,g)}\ .
\end{equation}
Next we integrate by parts in $\mathcal{D}g^z$. 
The relevant piece in the integration by parts is 
\begin{equation}
\int \mathcal{D}g^z\  e^{-S_g}\ \frac{\delta}{\delta g_z(t)}e^{i\Phi} = {\rm (bound.\ term)} + 
\int \mathcal{D}g^z\  e^{-S_g+i\Phi}\ \frac{\delta S_g}{\delta g_z(t)} = \int \mathcal{D}g^z\  e^{-S_g+i\Phi}\ 
\int dt'\ \frac{\Sigma^{-1}(t,t')}{\gamma(t)\gamma(t')}\ g^z(t')\ ,
\end{equation}
where the boundary term vanishes at $g^z(t)=\pm\infty$. 
So the integration by parts gives  
\begin{equation}
\begin{aligned}
\frac{E}{N} &= \int \mathcal{D}r\mathcal{D}s\mathcal{D}g^z\ 
e^{-S[g^z]+i\Phi(s,r,g)}\ \int dt dt'\ \gamma(t)\  \Sigma(t,t_f)\ s^z(t_f)\ \frac{\Sigma^{-1}(t,t')}{\gamma(t)\gamma(t')}\ g^z(t')
\ =\\
&=
\int \mathcal{D}r\mathcal{D}s\mathcal{D}g^z\ 
e^{-S[g^z]+i\Phi(s,r,g)} \frac{g^z(t_f)s^z(t_f)}{\gamma(t_f)}\ .
\end{aligned}
\end{equation}
Finally, integrating out $r$ we find 
\begin{equation}
\frac{E}{N} = \frac{\langle\langle g^z(t_f)s^z(t_f)\rangle\rangle}{\gamma(t_f)}\ ,
\end{equation}
where the average is over the self-consistent measure 
in Eq.~\eqref{eq:selfconsistentMeasure}. 
Note that since the noise covariance $\Sigma$ needs 
to be computed self-consistently, the effective field 
$g^z$ is not necessarily Gaussian, i.e. it is only 
Gaussian conditioned on the trajectory of the spins. 
If one succeeds to find the ground state, 
the energy should go to the Parisi value 
\begin{equation}
\lim_{t_f\to\infty}\frac{\langle\langle g^z(t_f)s^z(t_f)\rangle\rangle}{\gamma(t_f)}=-0.76321...\ .
\end{equation}

\subsection{Discrete map of the effective 
single-spin LLG dynamics}
\label{sec:effective-single-spin-LLG}

In this section we derive the discrete map corresponding 
to the effective dynamics
\begin{equation}
\frac{\partial s^{\alpha}}{\partial t} = \epsilon^{\alpha \beta \gamma} g^\beta(t) s^\gamma\ ,
\end{equation}
with $g^\beta(t)=(g^x(t),0,g^z(t))$ where 
$g^x(t)=\sum_n\beta_n\delta(t-n)$ and 
$g^z(t)$ is a stochastic field with 
\begin{equation}
\langle\langle g^z(t)g^z(t')\rangle\rangle =  
\gamma(t)\gamma(t')\ \langle\langle s^z(t)s^z(t')\rangle\rangle = 
\gamma(t)\gamma(t')\ \Sigma(t,t')
\end{equation}
with $\gamma(t) = \gamma_n$ constant in $(n-1, n)$. 
The dynamics evolves through $p$ layers, where 
a layer $n$ consists of two successive operations: 
first the local $z$-rotation driven by the continuous 
precession in the interval $t\in(n-1,n)$, and second 
the global $x$-rotation driven by the kick at the 
end of the interval $t=n$. The output of layer $n$ 
is the state of the spin after both rotations have 
occurred. 

\medskip

Now consider the continuous precession in the interval 
$t\in(n-1,n)$. The spin entering layer $n$ at time 
$t=(n-1)^+$ has just experienced the kick from layer 
$n-1$ so its state is $s(n-1)$. 
In the interval $t\in(n-1,n)$ the transverse field
$g^x=0$ and the stochastic field $g^z$ is active 
yielding the LLG equations  
\begin{equation}
\begin{aligned}
\dot{s}^x &= -g^z(t)\ s^y\ ,\\
\dot{s}^y &= +g^z(t)\ s^x\ ,\\
\dot{s}^z &= 0\ .
\end{aligned}
\end{equation}
The first thing to notice is that $s^z(t)$ is 
constant in the interval $t\in(n-1,n)$ and thus  
$s^z(t)=s^z(n-1)$. If we look at the covariance 
operator $\Sigma(t,t') =\langle\langle s^z(t)s^z(t')\rangle\rangle$ 
for $t\in(m-1,m)$ and $t'\in(n-1,n)$ we find that 
\begin{equation}
\Sigma(t,t') = \langle\langle s^z(m-1)s^z(n-1)\rangle\rangle\equiv
\Sigma_{mn} \qquad\ {\rm for}\
t\in(m-1,m),\ t'\in(n-1,n)\ ,
\label{eq:Sigma_mn}
\end{equation}
i.e., $\Sigma_{mn} = \Sigma(m-1,n-1)$ in the two time notation 
of the main text. 

We now look at the equations for the $x$ and $y$ 
components. Combining $s^x$ and $s^y$ into a single 
vector $w=(s^x,s^y)$ we obtain the differential equation 
$\dot{w}=g_z(t)\begin{pmatrix}
   0 &-1\\
   1 & 0
\end{pmatrix}w$, 
whose solution is 
\begin{equation}
w(t) = \begin{pmatrix}
   \cos(\theta_n) & -\sin(\theta_n)\\
   \sin(\theta_n) & \cos(\theta_n)
\end{pmatrix}
w(n-1) = R_z(\theta_n)w(n-1)\ ,
\end{equation}
where $\theta_n$ is the total $z$-rotation angle 
of the spin in layer $n$ given by 
\begin{equation}
\theta_n = \int_{n-1}^ndt\ g^z(t)\ .
\label{eq:theta_ndef}
\end{equation}
The angle $\theta_n$ is itself a Gaussian random 
variable (it is the integral of the Gaussian 
process $g^z(t)$) with zero mean and covariance 
\begin{equation}
\langle\langle \theta_m \theta_n\rangle\rangle = 
\int_{m-1}^mdt\ \int_{n-1}^ndt'\ 
\langle\langle g^z(t) g^z(t')\rangle\rangle = 
 \gamma_m\gamma_n
 \langle\langle s^z(m-1)s^z(n-1)\rangle\rangle = 
 \gamma_m\gamma_n\Sigma_{mn}\ ,
 \label{eq:thetacorrelation}
\end{equation}
where $\gamma_n$ is the coupling strength in layer $n$. 
Equation~\eqref{eq:thetacorrelation} is telling us 
that the matrix $\Sigma$ that dictates the angles 
$\theta_m$ and $\theta_n$ during layers $m$ and $n$ 
is defined by the spin states from the previous layers 
$m-1$ and $n-1$. This means that the discrete version 
of the dynamics does not require to solve the self-consistent 
fixed point condition in Eq.~\eqref{eq:self_consistency}. 
We clarify this important point layer by layer.

\subsubsection{{\bf Layer 1}}
In layer 1 we start by drawing an angle $\theta_1$ 
from a Gaussian distribution $\mathcal{N}(0,\gamma_1^2\Sigma_{11})$ 
so that 
\begin{equation}
\langle\langle \theta_1^2\rangle\rangle = \gamma_1^2\Sigma_{11} = 
\gamma_1^2 \langle\langle s^z(0)^2\rangle\rangle = 
\gamma_1^2\sin^2(\alpha)\ .
\end{equation}
As a consistency check with the single layer calculation 
performed in Sec.~\ref{sec:singleLayerSI} we now derive  
the optimal parameters $\alpha, \beta_1,\gamma_1$ 
that minimize the energy at time $t=1$
\begin{equation}
\frac{E}{N} = \frac{\langle\langle g^z(1)s^z(1)\rangle\rangle}{\gamma_1}\ ,
\end{equation}
where $g^z(1)$ is the stochastic field evaluated at 
time $t=1$. Now, we know that the field $g^z(t)$ is a 
continuous Gaussian process defined by its correlator 
$\langle\langle g^z(t)g^z(t')\rangle\rangle = 
\gamma(t)\gamma(t')\langle\langle s^z(t)s^z(t')\rangle\rangle$ 
and that $g^z(1)$ must be correlated with the angle 
$\theta_1$ via Eq.~\eqref{eq:theta_ndef}. More 
precisely 
\begin{equation}
\langle\langle g^z(1) \theta_1\rangle\rangle = 
\int_0^1dt\ \langle\langle g^z(1) g^z(t)\rangle\rangle = 
\gamma_1^2 \int_0^1dt\ \langle\langle s^z(1) s^z(t)\rangle\rangle = 
\gamma_1^2 \langle\langle s^z(1) s^z(0)\rangle\rangle\ ,
\end{equation}
where the last step follows from the fact that 
$s^z(t)=s^z(0)$ for $t\in(0,1)$. Using the expression 
for $s^z(0)$ and $s^z(1)$ 
\begin{equation}
\begin{aligned}
s^z(0) &= \sin(\alpha)\ ,\\
s^z(1) &= \sin(\beta_1)\cos(\alpha)\sin(\theta_1) + 
\cos(\beta_1)\sin(\alpha)\ ,
\end{aligned}    
\end{equation}
and taking the expectation 
we obtain
\begin{equation}
\langle\langle g^z(1) \theta_1\rangle\rangle = 
\gamma_1^2\ \cos(\beta_1)\sin^2(\alpha)\ .
\label{eq:g-theta-covariance}
\end{equation}
Therefore, $g^z(1)$ and $\theta_1$ are correlated 
Gaussian variables with zero mean and covariance 
given by Eq.~\eqref{eq:g-theta-covariance}. As 
a consequence we can express $g^z(1)$ as a linear 
combination of $\theta_1$ as 
\begin{equation}
g^z(1) = a\theta_1 + b\ ,
\end{equation}
where $b$ is an independent Gaussian variable 
and $a$ is given by 
\begin{equation}
a = \frac{\langle\langle g^z(1) \theta_1\rangle\rangle}{\langle\langle  \theta_1^2\rangle\rangle} = \cos(\beta_1)\ .
\end{equation}
The calculation of $\langle\langle g^z(1)s^z(1)\rangle\rangle$ 
thus reduces to 
\begin{equation}
\langle\langle g^z(1)s^z(1)\rangle\rangle = 
\cos(\beta_1)\langle\langle s^z(1) \theta_1\rangle\rangle = 
\cos(\beta_1)
\sin(\beta_1)\cos(\alpha)\gamma_1^2\sin^2(\alpha)\ 
e^{-\gamma_1^2\sin^2(\alpha)/2}\ .
\end{equation}
Defining $\tilde{\gamma}_1=\gamma_1\sin(\alpha)$, 
the energy function becomes 
\begin{equation}
\frac{E}{N} =
\sin(\alpha)\cos(\beta_1)\cos(\alpha)\sin(\beta_1)
\tilde{\gamma} e^{-\tilde{\gamma}^2 /2}
\end{equation}
which matches exactly the cost function in 
Eq.~\eqref{eq:cost1layer}.

\subsubsection{{\bf From Layer 2 to Layer $p$}}
A similar calculation can be performed for two layers 
to find the average energy as a function of the 
variational parameters $\alpha,\gamma_1,\beta_1,\gamma_2,\beta_2$. 
However, taking the gradient of the cost function 
results in a set of complicated equations with increasingly 
more terms as we go to higher layers. 
Instead of going in that direction, we look here for 
an efficient numerical procedure to compute the 
optimal parameters for an arbitrary number of layers by taking advantage of the purely forward structure of the covariance matrix $\Sigma_{mn}$. 

\medskip

The initial state of the spin at $t=0$ is $s^z(0)=\sin(\alpha)$. 
Therefore the entry $\Sigma_{11}$ of the covariance matrix 
is simply
\begin{equation}
\Sigma_{11} = \langle\langle s^z(0)^2\rangle\rangle=\sin^2(\alpha)\ .
\label{eq:1covariance}
\end{equation}

For $t\in(0,1)$ we need to generate the angle $\theta_1$. 
Since the covariance matrix of the angles contains the 
$\gamma$ factors, it is convenient to define 
\begin{equation}
C_{mn} = \gamma_m \gamma_n \Sigma_{mn}\ .
\end{equation}
We know that $\theta_1$ is a Gaussian random variable 
with zero mean and variance $C_{11}=\gamma_1^2\Sigma_{11}$, 
so it can be written as 
\begin{equation}
\theta_1 = \sqrt{C_{11}}\ x_1 =\gamma_1\sqrt{\Sigma_{11}}\ x_1\ , 
\qquad x_1\sim\mathcal{N}(0,1)\ .    
\end{equation}
We sample $\theta_1$ and we apply the $z$-rotation 
$R_z(\theta_1)$ and the kick $R_x(\beta_1)$ to get 
the new state of the spin $s(1)$
\begin{equation}
s(1) = R_x(\beta_1)R_z(\theta_1)s(0)\ ,
\end{equation}
whose $z$-component is explicitly given by 
\begin{equation}
s^z(1) = \sin(\beta_1)\cos(\alpha)\sin(\theta_1) + 
\cos(\beta_1)\sin(\alpha)\ .
\end{equation}
Note that $s^z(1)$ is also a random variable since 
it is a function of the random variable $\theta_1$ we 
just sampled. 

\medskip

For $t\in(1,2)$ we calculate the next entries of the 
covariance matrix 
\begin{equation}
\begin{aligned}
C_{12} &= \gamma_1\gamma_2\Sigma_{12} = 
\gamma_1\gamma_2\langle\langle s^z(0)s^z(1)\rangle\rangle\ ,\\
C_{22} &= \gamma_2^2\Sigma_{22} = 
\gamma_2^2\langle\langle s^z(1)^2\rangle\rangle\ .
\end{aligned}
\label{eq:2covariance}
\end{equation}
Next we have to sample $\theta_2$, but we cannot 
do it freely because we have already drawn a specific 
value for $\theta_1$, so we must sample $\theta_2$ 
conditioned on $\theta_1$. We look for an angle 
$\theta_2$ of the form 
\begin{equation}
\theta_2 = a \theta_1 + bx_2\ \qquad x_2\sim\mathcal{N}(0,1)\ ,
\end{equation}
with coefficients $a$ and $b$ determined by imposing 
that the covariance of $\theta_1$ and $\theta_2$ is 
given by Eqs.~\eqref{eq:1covariance} and~\eqref{eq:2covariance}, 
thus yielding 
\begin{equation}
\begin{aligned}
a &= \frac{C_{12}}{C_{11}}\ ,\\
b &= \sqrt{C_{22} - C_{12}^2/C_{11}}\ .
\end{aligned}
\end{equation}
To proceed to higher layers we first notice the 
following fact. Given the i.i.d Gaussian random 
variables $x_1,x_2\sim \mathcal{N}(0,1)$ we can 
write the correlated random variables $\theta_1$ 
and $\theta_2$ as linear combinations 
\begin{equation}
\begin{aligned}
\theta_1 &= A_{11}x_1\ ,\\
\theta_2 &= A_{21}x_1 + A_{22}x_2\ .
\end{aligned}
\end{equation}
So if we define the lower triangular matrix 
$A=\begin{pmatrix}
   A_{11} & 0\\
   A_{21} & A_{22}
\end{pmatrix}$ we can write 
\begin{equation}
C = \langle\langle \theta \theta^T\rangle\rangle =
\langle\langle Axx^TA^T\rangle\rangle = 
A\langle\langle xx^T\rangle\rangle A^T = AA^T\ ,
\end{equation}
which is just the Cholesky decomposition of $C$. 
The interesting point is that, if we add one 
more layer, the previous entries of the Cholesky 
matrix $A$ do not change as the matrix grows. 
Therefore we can iteratively compute the Cholesky 
factor $A$ one row at a time at each layer $n$. 
More precisely, at layer $n$, we calculate the 
covariance between the new angle $\theta_n$ and 
all previous angles $\theta_m$, for $m=1,...,n$ 
\begin{equation}
C_{mn} = \gamma_m\gamma_n\langle\langle s^z(m-1)s^z(n-1)\rangle\rangle\ .
\end{equation}
Then, find the new row in the Cholesky factor $A$ 
and compute the new angle $\theta_n$ as 
\begin{equation}
\theta_n = \sum_{i=1}^n A_{ni}\ x_i\ .
\end{equation}
Then apply the rotation $R_z(\theta_n)$ and the 
kick $R_x(\beta_n)$ and update the state of the 
spin as 
\begin{equation}
s(n) = R_x(\beta_n)\ R_z(\theta_n)\ s(n-1)\ .
\end{equation}
Then, repeat until the last layer $p$ to obtain 
the final state $s(p)$. 
At this point we need to compute the energy 
\begin{equation}
\frac{E}{N} = \frac{\langle\langle g^z(p)s^z(p)\rangle\rangle}{\gamma_p}\ . 
\end{equation}
The calculation of the average energy deserves 
some care, since the field $g^z(p)$ itself is 
correlated with all previous angles $\theta_1,...,\theta_p$. 
First of all we write $g^z(p)$ as a linear 
combination of the $\theta_n$'s as 
\begin{equation}
g^z(p) = \sum_{n=1}^pa_n\theta_n + b\ ,
\end{equation}
where $b$ is an independent Gaussian variable 
and the coefficient $a_n$ can be computed by 
taking averages of $g^z(p)$ with $\theta_m$ 
as 
\begin{equation}
\langle\langle g^z(p)\theta_m\rangle\rangle = 
\sum_{n=1}^pa_n \langle\langle \theta_n\theta_m\rangle\rangle = 
\sum_{n=1}^pa_n C_{nm}\ .
\end{equation}
Using the fact that 
\begin{equation}
\langle\langle g^z(p)\theta_m\rangle\rangle = 
\int_{m-1}^mdt  \langle\langle g^z(p)\ g^z(t)\rangle\rangle = 
\gamma_p\gamma_m\langle\langle s^z(p)\ s^z(m-1)\rangle\rangle \equiv u_m\ ,
\end{equation}
we can find the coefficients $a_n$ as 
\begin{equation}
a_n = \sum_{m=1}^p C^{-1}_{nm}\ u_m\ . 
\end{equation}
So the average $\langle\langle g^z(p)s^z(p)\rangle\rangle$ 
becomes 
\begin{equation}
\langle\langle g^z(p)s^z(p)\rangle\rangle = 
\sum_{n=1}^p
\sum_{m=1}^p C^{-1}_{nm}\ u_m 
\langle\langle \theta_n s^z(p)\rangle\rangle\ .
\end{equation}
Now the spin in the final state $s^z(p)$ is 
a function of all angles 
\begin{equation}
s^z(p) = f(\theta_1,...,\theta_p)\ ,  
\end{equation}
and we want to compute 
\begin{equation}
\langle\langle \theta_n s^z(p)\rangle\rangle = 
\int d\theta\ P(\theta)\ \theta_n\ f(\theta)\ .
\end{equation}
Using the identity 
\begin{equation}
\frac{\partial P(\theta)}{\partial\theta_k} = 
-P(\theta)\sum_{l=1}^p C^{-1}_{kl}\ \theta_l\ \to\ 
\sum_{k=1}^pC_{nk}\ \partial_kP(\theta) = -P(\theta)\ \theta_n
\end{equation}
we find that 
\begin{equation}
\int d\theta\ P(\theta)\ \theta_n\ f(\theta) = 
\sum_{k=1}^p C_{nk}\int d\theta\ P(\theta)\ 
\frac{\partial f(\theta)}{\partial\theta_k}\ ,
\end{equation}
and thus 
\begin{equation}
\langle\langle \theta_n s^z(p)\rangle\rangle = 
\sum_{k=1}^p C_{nk}\ \Big\langle\Big\langle 
\frac{\partial s^z(p)}{\partial \theta_k}\Big\rangle\Big\rangle
\end{equation}
Substituting in the equation for 
$\langle\langle g^z(p)s^z(p)\rangle\rangle$ 
and using the definition of $u_m$ we finally 
find the expression for the average energy 
\begin{equation}
\boxed{\ 
\frac{E}{N} = \sum_{k=1}^p\gamma_k\ 
\langle\langle s^z(p)\ s^z(k-1)\rangle\rangle\ 
\Big\langle\Big\langle 
\frac{\partial s^z(p)}{\partial \theta_k}\Big\rangle\Big\rangle\ }\ ,
\end{equation}
which is precisely Eq.~\eqref{eq:energydecomposition} 
in the main text.

\section{Effective single-spin description of QAOA on the SK model}
\label{sec:qaoaG}

The quantum quantities in Table~\ref{tab:viral-qaoa} 
and Fig.~\ref{fig:figure4} are computed from the exact 
infinite-size solution of the QAOA on the SK model of 
Farhi {\it et al.}~\cite{farhi2022}, in the $G$-matrix 
form of Ref.~\cite{boulebnane2025evidencequantumapproximateoptimization}. 
After averaging over the couplings, each site reduces 
to a single Ising variable $a_j$ on the $n=2p+1$ points 
of a closed time contour ($j=1,...,p$ the forward layers, 
$j=0$ the measurement time, and $j=-p,...,-1$ the backward 
layers), coupled through $\Gamma=(\gamma_{1},...,\gamma_{p}, 0, -\gamma_{p},...,-\gamma_{1})$. 
The contour two-point function $G_{jk}$ is a complex symmetric 
matrix ($G_{jk}=G_{kj}$) obtained by iterating 
\begin{equation}
\begin{aligned}
G^{(m)}_{jk} &= \sum_a f(a)\, a_ja_k\, 
\exp\Bigg[-\frac{1}{2}\sum_{rs} a_r \Gamma_r \Gamma_s\ G^{(m-1)}_{rs} a_s\Bigg]\ ,
\end{aligned}
\label{eq:Giter}
\end{equation}
where $f(a)$ is given by 
\begin{equation}
f(a) = \frac{1}{2}
\prod_{l=1}^{p}\langle a_l|e^{+i\beta_l X}|a_{l+1}\rangle
\prod_{l=1}^{p}\langle a_{-(l+1)}|e^{-i\beta_l X}|a_{-l}\rangle\ ,
\end{equation}
with the convention $a_{p+1}=a_{-(p+1)}=a_0$. The weight 
$f(a)$ is the product of the transition amplitudes of the 
qubit along the closed contour, where each link carries 
the mixer $e^{+i\beta_l X}$ on the forward branch and its 
conjugate $e^{-i\beta_l X}$ on the backward branch. The 
factor $1/2$ comes from the product of the two boundary 
overlaps $\langle a_1|+\rangle\langle +|a_{-1}\rangle = 1/2$, 
which is the same for every configuration. 

The iteration terminates exactly after $p$ updates~\cite{farhi2022, boulebnane2025evidencequantumapproximateoptimization}. 
The Gaussian kernel in Eq.~\eqref{eq:Giter} shows that each 
qubit is equivalent to a single spin in a Gaussian field of 
covariance $\Gamma_j \Gamma_k\ G_{jk}$, which is the 
exact quantum analog of Eq.~\eqref{eq:noise_def}. Furthermore, 
the constraint $Z_i^2=1$ pins the diagonal to $G_{jj}=1$. 
The energy density is given by 
\begin{equation}
\frac{E}{N} = \frac{i}{2}\sum_j \Gamma_j\, G_{0j}^2 = -\nu_p\ .
\label{eq:nup}
\end{equation}
Here we use the sign convention of Table~\ref{tab:viral-qaoa}, 
i.e. $\gamma_n<0$; with the angles of Ref.~\cite{basso22}, 
which have the opposite sign, the same expression in 
Eq.~\eqref{eq:nup} equals $+\nu_p$. 

Splitting the sum in the two branches of the contour 
and using the fact that $G_{0,-n}= G_{0n}^*$ yields
\begin{equation}
\frac{E}{N} = -2\sum_{n=1}^p\gamma_n\mathrm{Re}[G_{0n}]\, \mathrm{Im}[G_{0n}] = 
\sum_{n=1}^p\gamma_n C(n) R(n)\ ,
\end{equation}
with $C(n)=\mathrm{Re}[G_{0n}]$ and $R(n)=-2\mathrm{Im}[G_{0n}]$ 
as in Table~\ref{tab:viral-qaoa}, where $G_{0n}=G(n-1,p)$ 
in the two-time notation. The QAOA energy thus admits 
exactly the same layer decomposition as the VIRAL energy 
in Eq.~\eqref{eq:energydecomposition}. 

Each update is a direct sum over the $2^{2p+1}$ contour 
configurations, reduced  to $2^{2p}$ by the $\mathbb{Z}_2$ 
symmetry $a\to-a$, which is feasible up to $p=18$. The 
angles are the exact infinite-size optimized parameters 
of Basso {\it et al.}~\cite{basso22}, as distributed in 
the QOKit repository (the angles of Ref.~\cite{boulebnane2025evidencequantumapproximateoptimization}, 
obtained with matrix-product-state simulations at larger 
depths $p>20$, enter only in Fig.~\ref{fig:figure2}a). 
As a consistency check we verified that the normalization 
and the diagonal $G_{jj}=1$ hold to better than $10^{-8}$ 
at every update, and that Eq.~\eqref{eq:nup} reproduces the 
tabulated $\nu_p$ of Ref.~\cite{basso22} to better than $10^{-5}$. 

\label{LastBibItem}
\end{document}